\documentclass[twocolumn,resetfootnote]{aastex701}

\usepackage{amsmath}

\begin{document}

\title{Magnetically Driven Retrograde Precession in Misaligned Black Hole Accretion Flows}

\author[orcid=0000-0003-0292-2773]{Hong-Xuan Jiang}
\email{hongxuan\_jiang@sjtu.edu.cn}
\affiliation{Tsung-Dao Lee Institute, Shanghai Jiao Tong University, Shengrong Road 520, Shanghai, 201210, China}

\author[orcid=0000-0002-8131-6730]{Yosuke Mizuno}
\email{mizuno@sjtu.edu.cn}
\affiliation{Tsung-Dao Lee Institute, Shanghai Jiao Tong University, Shengrong Road 520, Shanghai, 201210, China}
\affiliation{School of Physics and Astronomy, Shanghai Jiao Tong University, 
800 Dongchuan Road, Shanghai, 200240, China}
\affiliation{Key Laboratory for Particle Physics, Astrophysics and Cosmology, Shanghai Key Laboratory for Particle Physics and Cosmology, Shanghai Jiao-Tong University, 800 Dongchuan Road, Shanghai, 200240, China}
\affiliation{Institut f\"ur Theoretische Physik, Goethe-Universit\"at Frankfurt, Max-von-Laue-Stra{\ss}e 1, D-60438 Frankfurt am Main, Germany}

\author{Dong Lai}
\email{donglai@sjtu.edu.cn}
\affiliation{Tsung-Dao Lee Institute, Shanghai Jiao Tong University, Shengrong Road 520, Shanghai, 201210, China}
\affiliation{Center for Astrophysics and Planetary Science, Department of Astronomy, Cornell University, Ithaca, NY 14853, USA}

\author[orcid=0000-0002-4064-0446]{Indu K. Dihingia}
\email{ikd4638@gmail.com}
\affiliation{Tsung-Dao Lee Institute, Shanghai Jiao Tong University, Shengrong Road 520, Shanghai, 201210, China}

\author{Christian M. Fromm}
\email{christian.fromm@uni-wuerzburg.de}
\affiliation{Institut f\"ur Theoretische Physik und Astrophysik, Universit\"at W\"urzburg, Emil-Fischer-Str. 31, D-97074 W\"urzburg, Germany}
\affiliation{Institut f\"ur Theoretische Physik, Goethe-Universit\"at Frankfurt, Max-von-Laue-Stra{\ss}e 1, D-60438 Frankfurt am Main, Germany}
\affiliation{Max-Planck-Institut f\"ur Radioastronomie, Auf dem H\"ugel 69, D-53121 Bonn, Germany}

\correspondingauthor{Hong-Xuan Jiang, Yosuke Mizuno}
\email{hongxuan\_jiang@sjtu.edu.cn, mizuno@sjtu.edu.cn}

\begin{abstract}
Observations of accreting black hole (BH) systems, such as microquasars and supermassive black holes, often reveal a precessing jet with changing directions, indicating a misaligned accretion flow relative to the BH spin.
The precession is commonly attributed to the Lense-Thirring (LT) effect, which arises from the BH's rotation twisting the surrounding spacetime and accretion flow.
In the strongly magnetized regime, which is preferred accretion flow conditions for M~87$^*$ and likely other jet-producing systems, the large-scale magnetic field can significantly influence the flow dynamics. Here, we perform large-scale three-dimensional general relativistic magnetohydrodynamic simulations of tilted accretion onto a rotating BH, and find a never-seen-before new retrograde precession. This precession arises from a magnetic torque on the disk generated by the poloidal magnetic field aligned with the BH's rotation, opposing the LT torque. This finding highlights the unique property of highly magnetized accretion flows around BHs and provides a new interpretation of jet precession observed in many systems.
\end{abstract}



\section{Introduction} \label{sec:intro}
Black holes (BHs), especially supermassive black holes in the center of galaxies, are widely believed to be rotating Kerr BHs \citep{2002ApJ...565L..75E}. 
Similarly, stellar-mass black holes in X-ray binary systems, known as microquasars, often exhibit rapid rotation ($a \gtrsim 0.9$), as measured from observational constraints \citep{2024ApJ...967L...9Z, 2011ApJ...742...85G, 2011MNRAS.416..941S, 2020A&A...643A..31K}. In these systems, the BH spin and orbital angular momentum are often found to be misaligned which introduces significant complexity to the accretion and jet dynamics
 \citep{2010ApJ...719L..79F, 2022Sci...375..874P}.
The rotation of a BH twists the surrounding spacetime through frame-dragging, leading to the effect known as the Lense-Thirring (LT) precession \citep{1918PhyZ...19...33T}. This rotation introduces unique relativistic phenomena that significantly influence the surrounding accretion flow and the configuration of magnetic fields \citep{Narayan2003}. 

Typically a BH's spin magnitude and direction are determined by its formation and cumulative accretion history \citep{2021ARA&A..59..117R, 2020A&A...635A..97B}. However, the angular momentum of the infalling plasma can often have a different orientations, resulting in spin-disk misalignment and the formation of tilted or warped accretion flow structures \citep{2005ApJ...623..347F, 2007ApJ...668..417F, Liska2018, 2021MNRAS.507..983L, Chatterjee2023}. 
Several microquasars present evidence of jet precession, such as SS 433, GRO J1655-40, and GRS 1758-258 \citep{1984ARAA..22..507M,2014MNRAS.437.2554M,2015A&A...584A.122L, 2002ApJ...578L.129S}.
Misalignments of accretion flows are naturally expected in tidal disruption events (TDEs), since the impact direction of the star is purely random \citep{2019MNRAS.487.4965Z,2014MNRAS.437.2744T,2016MNRAS.455.1946F,2024Natur.630..325P, 2012PhRvL.108f1302S, Andalman2022}. Similarly, the major merger events of active galactic nuclei (AGN) can also cause misalignment between the rotation axis of the remnant BH and accreting plasma. Several observations have revealed evidence of tilted accretion disks in the AGN systems \citep[e.g., ][]{2005ApJ...618..618K, 2006ApJ...638..120C, 2007MNRAS.379..135C}. In the case of geometrically thin accretion flows, disk warps propagate diffusively due to viscosity, the combination of differential LT precession and viscous dissipation gives rise to the Bardeen-Petterson effect \citep{1975ApJ...195L..65B}, in which tends to align with the BH's spin axis (see \cite{1996MNRAS.282..291S,2019MNRAS.487..550L, 2021MNRAS.507..983L}). In geometrically thick accretion flows, the disk warp propagates as bending waves, LT precession can occur in Standard and Normal Evolution (SANE) flows \citep{2005ApJ...623..347F, 2007ApJ...668..417F, 2008ApJ...687..757F, Liska2018, 2019ApJ...878...51W, 2019MNRAS.487.4965Z}. From simulations, global precession only happens when the sound-crossing time within the accretion flow is shorter than the precession timescale \citep{2024arXiv240410052F}. The magnetorotational instability (MRI) drives turbulence and angular momentum transport in accretion disks, causing the disk to expand radially, while the redistribution of angular momentum alters the precession rate, generally slowing it down over time \citep{Liska2018}. This suggests that the torus size may play a key role in facilitating the LT precession of both the disk and the jet.

Simulations of tilted MAD suggest that powerful jets can drive the disk to align with the BH’s equatorial plane \citep{McKinney2013, 2017MNRAS.464.2660P, 2021MNRAS.504.6076R, Chatterjee2023}. \cite{Chatterjee2023} carried out tilted MAD simulations with a widely used large torus model extending beyond $800\,r_{\rm g}$ (where $r_{\rm g}=GM/c^2$ is the gravitational radius) found, no clear evidence of disk or jet precession \citep{McKinney2013, Chatterjee2023}. Thus, there are uncertainties to date about whether precession happens in MAD flow or not. This is an important question to study further, as MAD accretion flows are strongly favored in the case of M~87$^*$ and likely other jet producing systems \citep{Akiyama2019, Yuan2022}. Indeed, long-term observation of M~87$^*$ has revealed strong evidence of jet precession \citep{Cui2023}. This suggests that MAD flow and jet precession can coexist. In this work, we perform GRMHD simulations of tilted disks across various magnetic configurations, focusing on the precession of tilted MADs. For the first time, we find that MAD with sufficiently large magnetic flux can undergo magnetically driven retrograde precession, overwhelming the prograde LT precession driven by the BH spin. Our simulation demonstrate that it is possible to generate precession of jet and disk in MADs and provide a new perspective to the current observations.

The paper is organized as follows: we present our results in Section 2 and conclude in Section 3; code setup and supplementary information can be found in the Appendix.
\section{Results}\label{sec2}

\begin{figure*}
\centering 	
\includegraphics[width=\linewidth]{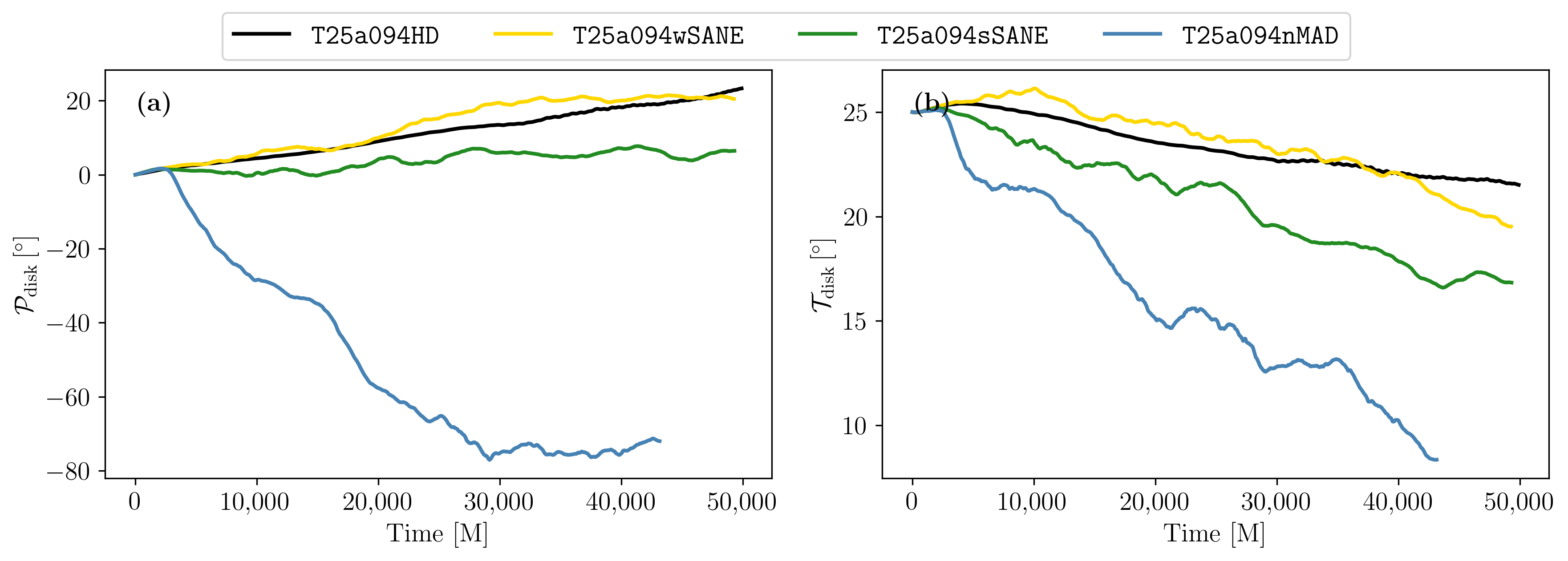}
\caption{Panels (a) and (b) are the time evolution of disk precession and tilt angles of the models {\tt T25a094HD} (black), {\tt T25a094wSANE} (yellow), {\tt T25a094sSANE} (green), and {\tt T25a094nMAD} (blue). Details on the notion of the models can be found in Table.~\ref{table:models}.}
 	\label{fig: precession_exmple}
\end{figure*}

\begin{figure*}
\centering 	
\includegraphics[width=.49\linewidth]{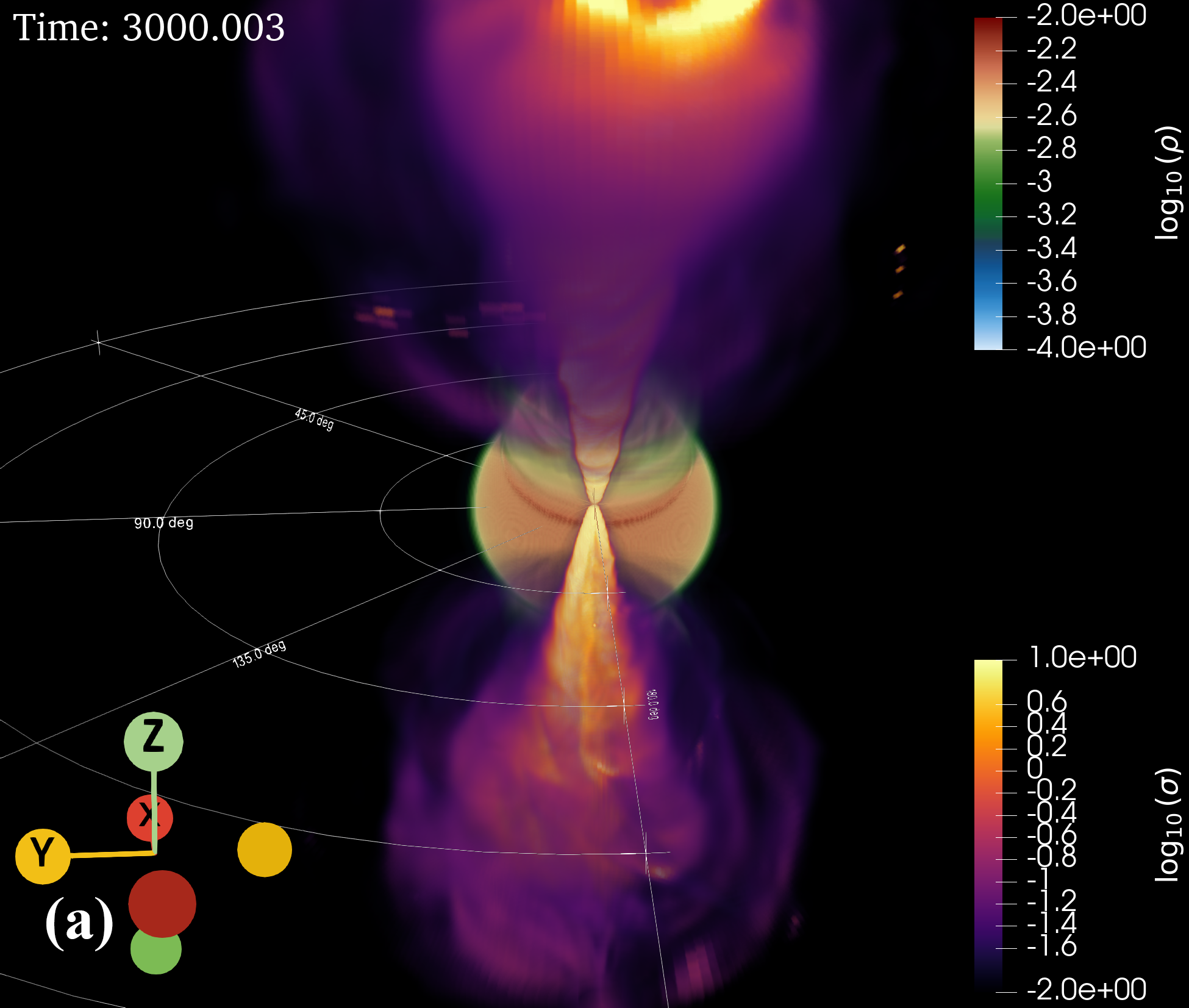}
\includegraphics[width=.49\linewidth]{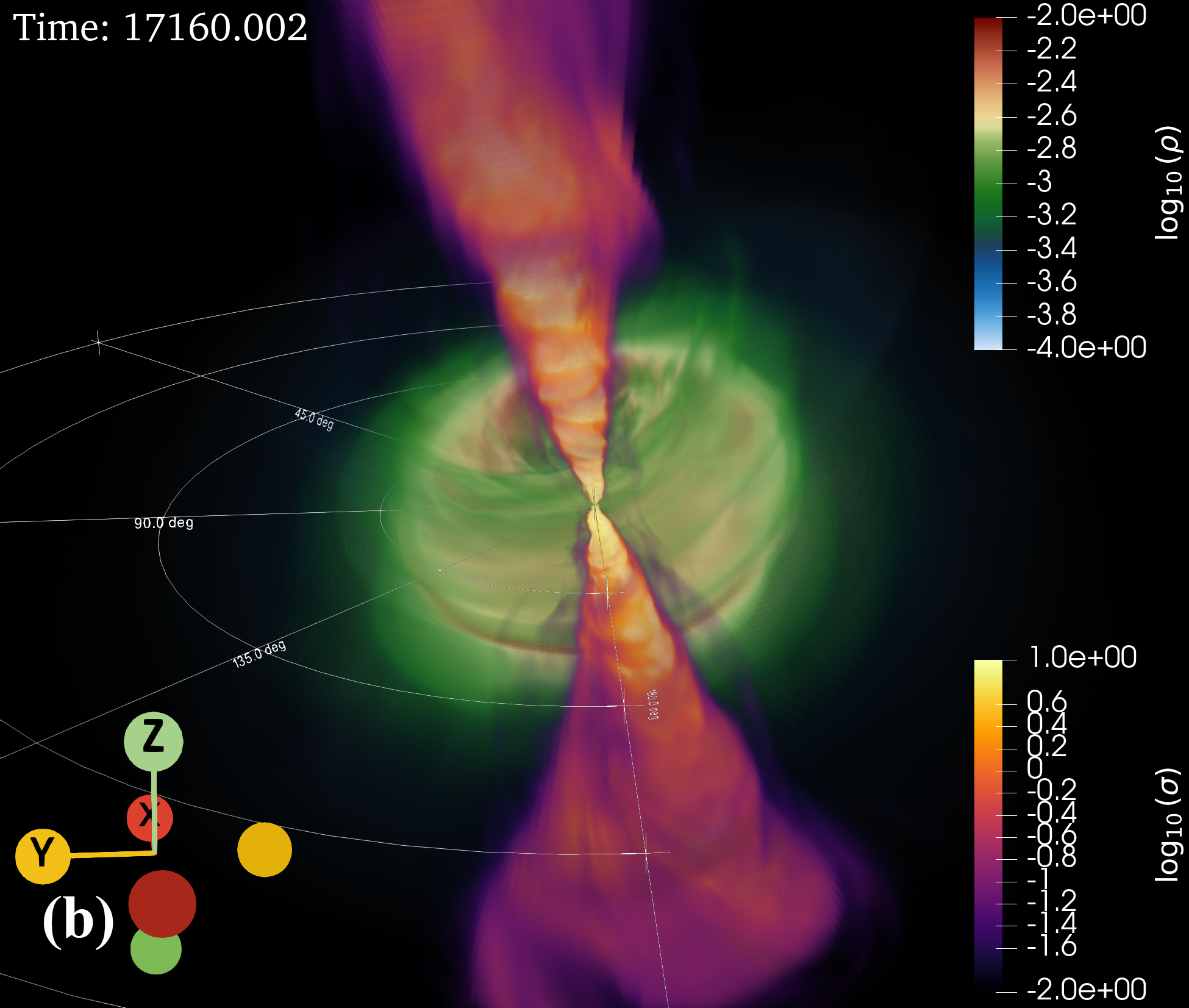}
 	\caption{3D Volume-rendering of the jet (colored by magnetization $\sigma$) and accretion disk (colored by density $\rho$) in the strongly-magnetized model~\texttt{T25a094nMADH}, shown at two different time snapshots: 
        (a)~$t = 3{,}000\,\rm M$ and (b)~$t = 17{,}160\,\rm M$. 
        The retrograde precession of the jet-disk system is visible as a counter-rotation to the BH spin axis (positive $z-$direction).}
    \label{fig: volume_rendering}
\end{figure*}

\begin{figure*}
\centering 	
\includegraphics[width=\linewidth]{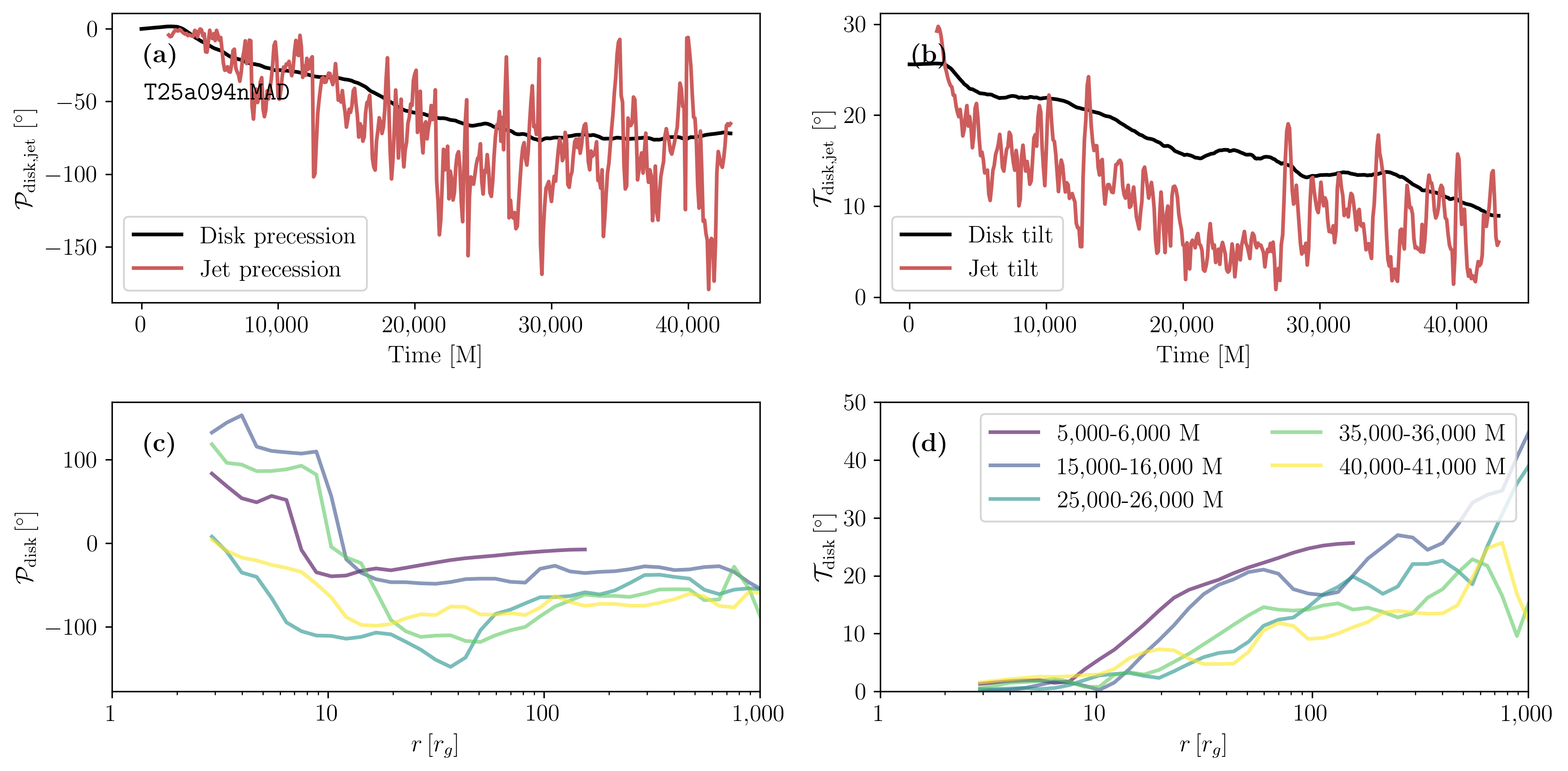}
\caption{Evolution and radial profiles of jet and disk precession and tilt angles.
Panels (a) and (b) depict the time evolution of the precession and tilt angles for the disk (black) and the jet (red) for the strongly-magnetized model {\tt T25a094nMAD}. Panels (c) and (d) show the radial profiles of the disk’s precession and tilt angles for the model {\tt T25a094nMAD}, with different colors indicating different averaging time ranges.}
 	\label{fig: T25a094nMAD}
\end{figure*}

Here, we perform GR(M)HD simulations of tilted disks using KHARMA, a GPU-accelerated extension of the {\tt iharm3D} code \citep{2021JOSS....6.3336P}. To explore the interaction between magnetic fields and the LT effect, we conduct a series of simulations varying the magnetic field configurations and the BH spin $a$.
Detailed numerical methods and resolution information are presented in Appendix~\ref{sec: numerical_method} and \ref{sec: resolution}. 

\subsection{Precession and tilt from GRMHD simulations}

\begin{figure*}
\centering 	
\includegraphics[height=.41\linewidth]{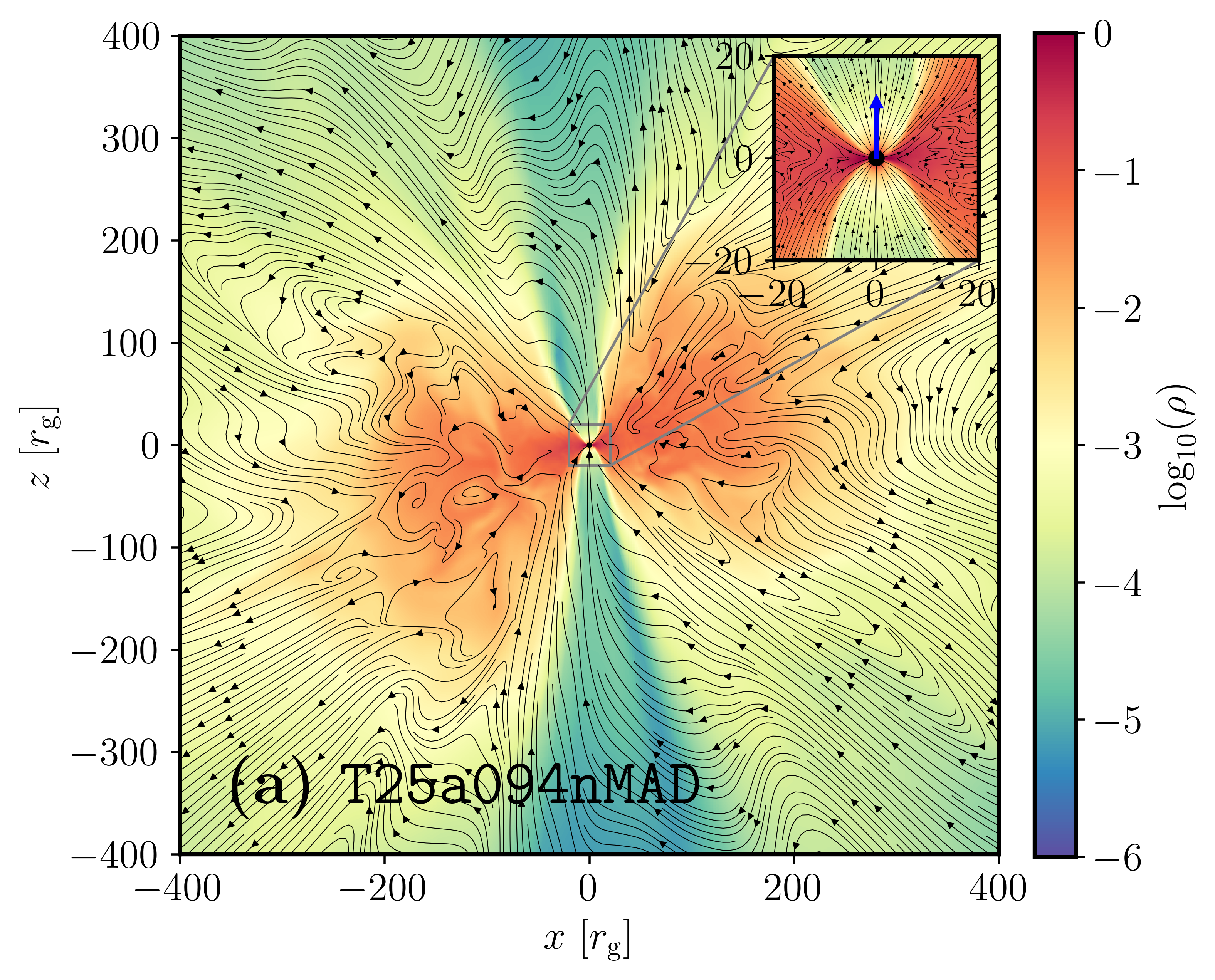}
\includegraphics[height=.41\linewidth]{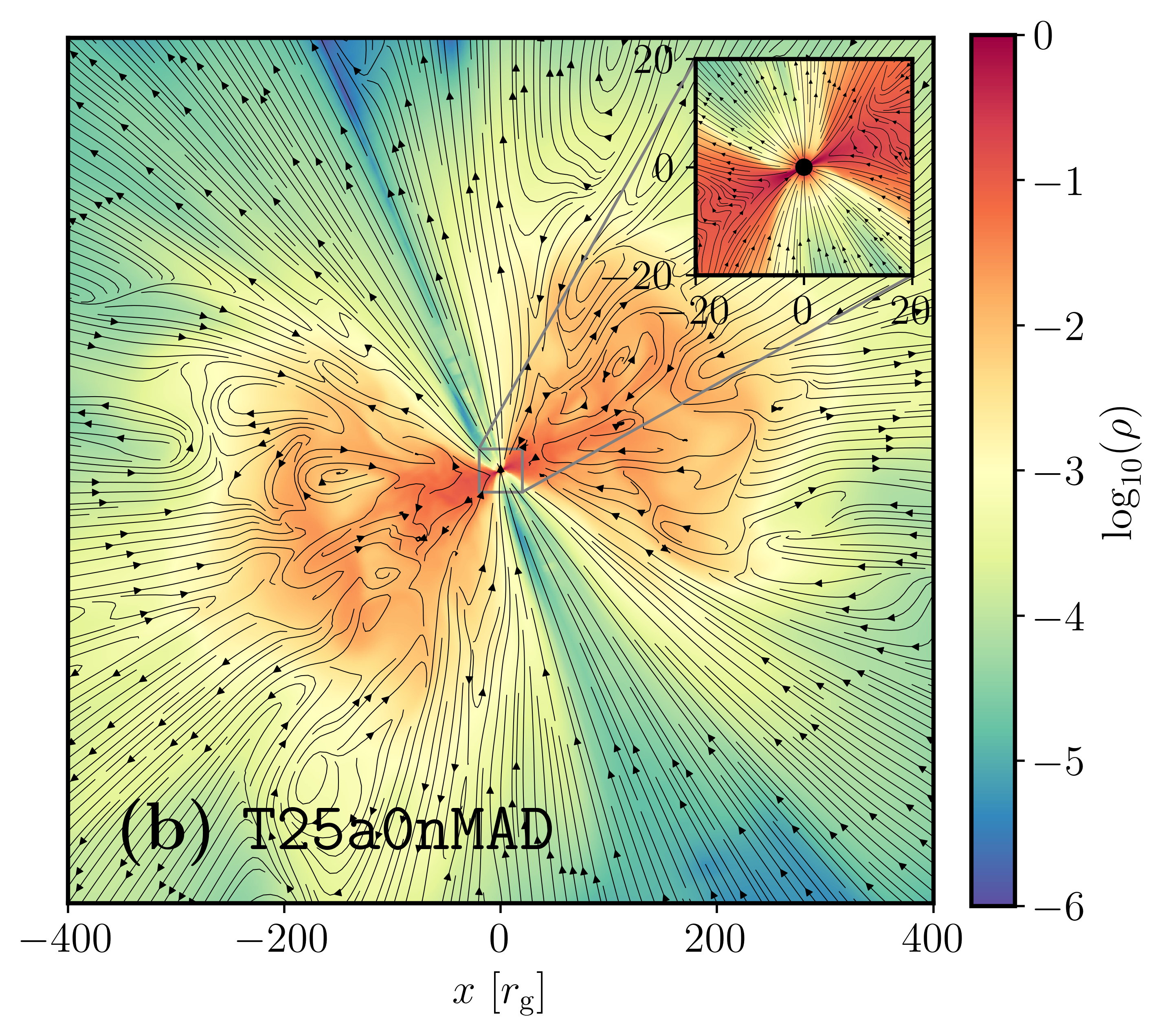}
 	\caption{Distributions of time-averaged logarithmic density on the $x-z$ plane of the rapidly-rotating BH model {\tt T25a094nMAD} and non-rotating BH model {\tt T25a0nMAD} with $a=0.9375$ and $a=0$ in panels (a) and (b), respectively. The streamlines in each panel are the time-averaged poloidal magnetic field lines. The time averaging range is from $t=14,000\,\rm M$ to $15,000\,\rm M$. The blue arrow in the zoomed-in part of panel (a) shows the BH spin direction.}
    \label{fig: stream}
\end{figure*}

In Fig.~\ref{fig: precession_exmple}, we present the evolutions of the disk precession angle ($\mathcal{P}_{\rm disk}$) and tilt angle ($\mathcal{T}_{\rm disk}$) for four representative models. The definitions of the disk and jet precession and tilt angles follow \cite{Liska2018} and are detailed in Appendix\footnote{The tilt angle is defined as the angle between the disk angular momentum and the positive $z$-axis, while the precession angle corresponds to the position angle of the disk angular momentum. The jet orientation follows \cite{Liska2018} and is determined by isolating the jet region based on magnetic pressure and computing its tilt and position angle similarly to the disk.}.

In the magnetically-weak SANE (yellow and green lines) and non-magnetized HD (black line) models, the LT effect leads to prograde precession, following the direction of the BH spin. 
Model {\tt T25a094HD} exhibits nearly persistent precession with approximately constant disk size. 
In the two SANE models ({\tt T25a094wSANE} and {\tt T25a094sSANE}) with different magnetic field strengthes, the time evolutions of precession and tilt angles show similar trends and magnitudes that is comparable values presented in \cite{Liska2018}. 
The most weakly magnetized model {\tt T25a094wSANE} closely follows the corresponding non-magnetized HD model {\tt T25a094HD}, indicating a minimal effect from the magnetic field. In contrast, the mildly magnetized SANE model {\tt T25a094sSANE} shows a slower precession rate and faster tilt alignment, driven by enhanced magneto-spin alignment. In both models, precession almost stops in later simulation time ($30,000$ and $40,000\,\rm M$).

However, once the magnetic field strength reaches the MAD threshold ($\Phi_{\mathrm{B}}/\sqrt{\dot{M}} \sim 15$, for co-rotating BHs \citep{2011MNRAS.418L..79T}), the disk precession reverses its direction. Shown by the blue line in Fig.~\ref{fig: precession_exmple}(a) for the highly-magnetized model {\tt T25a094nMHD}, the disk initially undergoes prograde precession before jet launch due to the LT effect, overlapping with the non-magnetized model {\tt T25a094HD}. 
Remarkably, as the accretion flow plunges into the BH and the jet launches, we observe rapid retrograde precession opposing the BH rotation.
This indicates that in highly magnetized accretion flows, the precession is mainly governed by magnetic forces. The resulting magnetic torque counteracts the LT torque produced by the spinning BH.
This retrograde precession is also evident in the volume-rendering images of the highly magnetized model~{\tt T25a094nMAD}, shown in Fig.~\ref{fig: volume_rendering}. In this model, the BH spin is aligned with the $\hat{z}$ direction, while the disk and jet are initially tilted by $25^\circ$. A comparison between panels (a) and (b) of Fig.~\ref{fig: volume_rendering} reveals a clear shift in the orientation of the jet and disk from $t = 4,700\,M$ to $15,000\,M$, precessing in a direction opposite to the BH rotation.
Meanwhile, in all the magnetized models, the tilt angle continuously decreases due to magneto-spin alignment \citep{McKinney2013, Chatterjee2023}, as shown in Fig.~\ref{fig: precession_exmple}(b). Since the tilted disk condition is not an equilibrium state exactly, the tilt angle in model {\tt T25a094HD} gradually decreases over time.

Precession and alignment occur not only in the disk but also in the jet. As shown in Fig.~\ref{fig: T25a094nMAD}(a) and (b), both the jet and disk in highly-magnetized simulation {\tt T25a094nMAD} exhibit evolving precession and tilt angles. While the jet's precession generally tracks the disk's motion, it displays much stronger variability due to magnetic flux eruptions in MAD regime. Notably, the strong magnetic field in jet enhances magneto-spin alignment, resulting in a more rapid decrease of its tilt angle compared to the disk.

The reversed precession direction from prograde to retrograde in highly-magnetized simulation {\tt T25a094nMAD} indicates a dominance of magnetic torque opposing the LT torque. This torque balance between the magnetic and LT torque directly determines whether disk precession proceeds in a prograde or retrograde direction. Our results further suggest that the slower precession observed in the moderately-magnetized SANE model {\tt T25a094sSANE} may not be explained solely by disk expansion. Additionally, the stronger magnetic torque likely mitigates LT precession, causing it to slow down.

In the radial direction, the LT torque $\tau_{\rm LT}\propto r^{-3}$ decreases rapidly as radius increases. It suggests a strong LT precession torque in the innermost region of the disk. In Fig.~\ref{fig: T25a094nMAD}(c) and (d), we present the disk precession and tilt angle profiles of model {\tt T25a094nMAD} at different simulation times. The profiles are computed in the radial direction by dividing the disk into concentric rings, each spanning 8 grid cells in radius, and integrating the total angular momentum within each ring. The figure shows substantial radial variation in the disk precession angle. In the region $r \lesssim 10\,r_{\rm g}$, the disk orientation twists in the positive $\phi$-direction due to the frame-dragging effect of the rotating BH. Over time, however, the difference in the precession angle between the innermost and outer regions diminishes, as illustrated by the yellow line in Fig.~\ref{fig: T25a094nMAD}(c). After the precession stops around $t \sim 30,000\,\rm M$, the radial profile of the precession angle continues to evolve, gradually flattening out the difference between the innermost and outer regions.
In Fig.~\ref{fig: T25a094nMAD}(b), the disk tilt angle decreases in the inner region ($r \lesssim 10 \, \rm{r}_{\rm g}$), indicating a strong magneto-spin alignment, consistent with the findings of \cite{Chatterjee2023}. Consequently, in a strongly magnetized tilted disk, the inner and outer regions exhibit pronounced twisting and warping driven by the combined effects of LT effect, magnetic torques, and magneto-spin alignment (see Fig.~\ref{fig: stream}(a) for instance).

\begin{figure*}
\centering 	
\includegraphics[width=.8\linewidth]{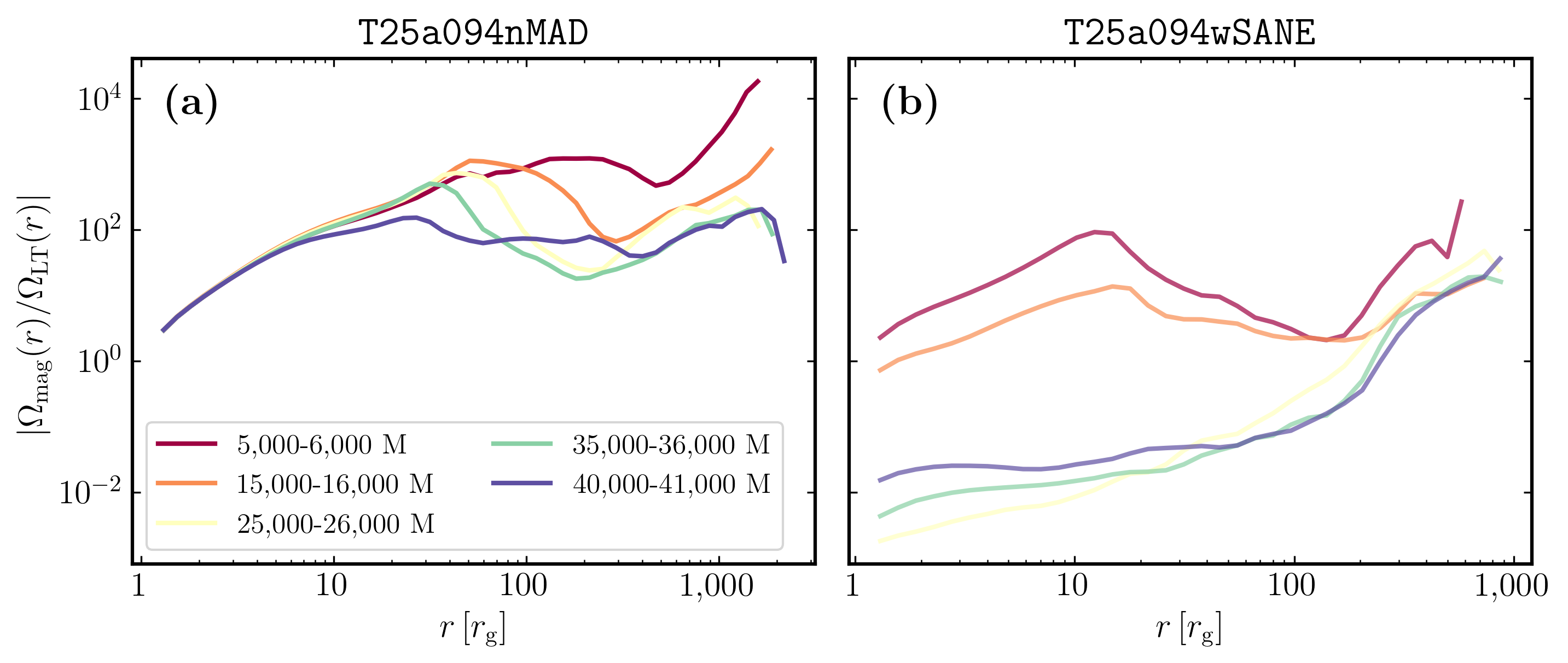}
\caption{This figure presents the radial profiles of the ratio of the magnetic to LT precession rates, $|\Omega_{\rm mag}/\Omega_{\rm LT}|$, for simulations (a) $\tt T25a094nMAD$ (left) and (b) $\tt T25a094wSANE$ (right). Each colored line depicts the time-averaged profile over the interval indicated in the legend. These profiles are calculated from the simulation data by averaging values within radially binned shells, where each bin spans 8 cells in the radial direction.}
 	\label{fig: Omega_ratio}
\end{figure*}

\subsection{Magnetic torque}

A natural question arises: if the angular momentum of the BH and disk is roughly the same direction, with only a small initial tilt angle between them, why does the magnetic field drive precession in the opposite direction? To understand this, we plot the averaged density distribution and poloidal magnetic field lines for the models {\tt T25a094nMAD} and {\tt T25a0nMAD} in Fig.~\ref{fig: stream} with an average range of $t=14,000-15,000\,\rm M$. Both simulations share identical torus setups, with the only difference being the BH spin: rapidly rotating $a = 0.9375$ and non-rotating $a = 0$ in panels (a) and (b), respectively. 
In the case of a non-rotating BH, where the frame-dragging effect is absent, the poloidal magnetic field lines remain predominantly radial. 
In contrast, for a rapidly-rotating Kerr BH (panel (a)), the poloidal magnetic field lines tend to align with the BH spin direction. 
This alignment suggests that the BH’s rotation drags the initially
tilted poloidal magnetic field. This dragging effect simultaneously
strengthens the toroidal component of the magnetic field and reorients
the poloidal field and the overall magnetic configuration to align
with the BH spin. On the other hand, the alignment of the inner disk
is substantially slower than that of the magnetic field. This implies
that the disk (angular momentum) axis and the magnetic field symmetry
axis can be misaligned for an appreciable period of time.

A simple model proposed by \cite{Lai2003} suggests that a misaligned disk in a vertical magnetic field $B_{z}$ (aligned with the BH spin axis $\hat{z}$) experiences a magnetic torque that induces retrograde precession. This precession torque exists whenever a conducting disk is embedded in an external, inclined magnetic field (see \cite{1999ApJ...524.1030L}). When the ``external" magnetic field $\mathbf{B}=B_z {\hat{z}}$ is misaligned with the disk axis $\hat{l}$, it projects a radial component in the disk plane, $B_r=B_z\sin\tilde{\beta}\sin{\phi}$ (where $\tilde{\beta}$ is the angle between $\hat{z}$ and ${\hat{l}}$, $\phi$ is the azimuthal angle around the disk). On the other hand, the perpendicular field $B_z\cos\tilde{\beta}\simeq B_z$ (for small $\tilde{\beta}$) induces an azimuthal screening surface current $K_\phi=(c/2\pi)B_z\tan{\theta}$ in the disk, where $\tan{\theta} = B_r^{\rm (ind)}/B_z$ and $B_r^{(\rm ind)}$ is the induced radial field on the upper disk surface. The interaction between $K_\phi$ and $B_r$ produces a $\phi-$dependent perpendicular force on the disk, leading to a magnetic torque (per unit area) on the disk \citep{Lai2003}:
\begin{equation}
    \mathbf{T_{\rm mag}}=-\frac{1}{4\pi}r B_z^2\tan{\theta}\hat{z}\times\hat{l}.
\end{equation}
This torque induces a (local) retrograde precession,
\begin{equation}
    \Omega_{\rm mag}(r) = -\frac{B_z^2\tan\theta}{4\pi\Sigma r \Omega(r)},
\end{equation}
where $\Sigma$ is the surface density of the disk, $\Omega(r)$ is the disk rotation (angular) frequency. Comparing with the LT precession frequency (with $G=c=1$, $S$ is the angular momentum of the BH)
\begin{equation}
    \Omega_{\rm LT}(r)=\frac{2S}{r^3}=\frac{2M^2a}{r^3},
\end{equation}
we find 
\begin{equation}
    \frac{\Omega_{\rm mag}(r)}{\Omega_{\rm LT}(r)}\simeq - \frac{B_z^2\tan\theta/8\pi}{a\rho c c_{\rm s} (r_{\rm g}/r)^2}, \label{Eq: ratio}
\end{equation} 
where we have used $\Sigma\sim\rho H\sim\rho c_{\rm s}/\Omega$, and $c_{\rm s}$ is the disk sound speed.
When the ratio $|\Omega_{\rm mag}/\Omega_{\rm LT}|$ is greater than 1, the magnetic torque dominates the LT torque.

Since the LT torque rapidly decreases as $r^{-3}$, angular momentum transport plays a crucial role in enabling global precession of the entire disk rather than just the innermost region \citep{Liska2018, 2024arXiv240410052F}. To quantitatively evaluate the contribution of magnetic torques in our GRMHD simulations (including both SANE and MAD regimes), we compute the radial profiles of the ratio
$|{\Omega_{\rm mag}}/{\Omega_{\rm LT}}|$ from our simulation data. Significant variation in the $|{\Omega_{\rm mag}}/{\Omega_{\rm LT}}|$ ratio is observed across MAD and SANE models. Fig.~\ref{fig: Omega_ratio} illustrates this difference, showing time-averaged profiles of the ratio for representative models: {\tt T25a094wSANE} (SANE) and {\tt T25a094nMAD} (MAD).
In the weakly-magnetized SANE model {\tt T25a094wSANE}, for $r \lesssim 100\,r_{\rm g}$, the ratio $|{\Omega_{\rm mag}} / {\Omega_{\rm LT}}|$ drops to the range $\sim 10^{-2}-10^{0}$ at late times ($t\gtrsim20,000\,\rm M$). This demonstrates the dominance effect from LT torque over magnetic torque, leading to global prograde precession in this model. In contrast, the magnetically-dominated MAD simulation ({\tt T25094nMAD}) shows $\langle \Omega_{\rm mag}/ \Omega_{\rm LT} \rangle \sim 10$, resulting in a much stronger magnetic torque contribution compared to model {\tt T25a094wSANE}, and leading to global retrograde precession.

\section{Discussions and conclusions}

In this work, we study the mechanism of magnetically driven retrograde precession in geometrically thick accretion disk systems. 
Our GRMHD simulations of tilted disk systems exhibit precession influenced by both the magnetic field and the LT effect. Notably, the magnetic torque tends to compensate for the LT torque. In strongly magnetized accretion flows, the magnetic torque dominates, resulting in retrograde precession, whereas in weakly magnetized flows, LT-driven prograde precession dominates. 
The precession rate strongly depends on the size of the torus, with both retrograde and prograde precession gradually slowing down and eventually ceasing as the disk expands. This behavior is consistent with the findings by \cite{Liska2018}.

In our simulations, the disk reaches a size of approximately $200\,r_{\rm g}$  which is smaller than the typical torus size for MAD simulations ($\sim 400\,r_{\rm g}$) \cite{event_horizon_telescope_collaboration_first_2022}. Without a magnetic field, the LT precession rate remains significantly lower than the M~87 jet precession rate reported in \cite{Cui2023}. This suggests that if the jet of M~87$^*$ is indeed precessing, an alternative, stronger torque is needed. Given the strong evidence that the accretion flow in M~87$^*$ is likely in the MAD regime \citep[e.g.,][]{Akiyama2019, Yuan2022}, we propose that the precession of the M~87$^*$ jet is more plausibly explained by magnetically driven retrograde precession rather than prograde LT precession. 
Our findings suggest that precession in jet-launching systems is not solely governed by the LT effect but is also significantly influenced by magnetic torques. 

Determining the direction of jet precession (prograde or retrograde) is key to understanding the underlying physics, but is observationally challenging. Often, a side-on view of the jet's projection onto the sky makes it difficult to definitively ascertain the sense of precession. However, there are opportunities. The clear circular motion in V404 Cygni's jet \cite{2019Natur.569..374M} reveals its precession direction; knowing the black hole spin orientation would then help determining whether the precession is retrograde or prograde. On the other hand, for jets such as M~87, variations in jet width during precession offer a path to discern the precession direction. These measurements will be vital for distinguishing between the magnetically-driven retrograde precession, and LT prograde precession, offering deeper insights into accretion dynamics and the role of magnetic fields near black holes, such observations will be feasible with future arrays such as the ngEHT and the ngVLA.

\begin{acknowledgments}

We acknowledge the Zhejiang Lab for providing computational facilities that contributed to this work and thank Hongzhe Zhou and Zhen Pan for useful discussions. 
This research is supported by the National Key R\&D Program of China (Grant No.\,2023YFE0101200), the National Natural Science Foundation of China (Grant No.\,12273022, 12192220, 12133008), and the Shanghai Municipality Orientation Program of Basic Research for International Scientists (Grant No.\,22JC1410600). The simulations were performed on TDLI-Astro, Pi2.0, and Siyuan Mark-I at Shanghai Jiao Tong University. 
\end{acknowledgments}

\software{KHARMA}

\appendix
\section{GRMHD code and models}\label{sec: numerical_method}

In this work, all GR(M)HD simulations are performed by the KHARMA\footnote{\url https://github.com/AFD-Illinois/kharma} code \citep{2024arXiv240801361P}, which is a GPU-accelerated version of {\tt iharm3D} \cite{2021JOSS....6.3336P}. Both KHARMA and {\tt iharm3D} originate from HARM \cite{2003ApJ...589..444G}, which solves the ideal MHD equations in the framework of general relativity. The ideal GRMHD equations are solved for an Eulerian observer and are written as follows:
\begin{equation}
    \begin{aligned}
        \partial_t(\sqrt{-g}\rho u^t)&=-\partial_i(\sqrt{-g}\rho u^i),\\
        \partial_t(\sqrt{-g}T^t_{\nu})&=-\partial_i(\sqrt{-g} T^i_{\nu}) + \sqrt{-g}T^\kappa_\lambda\Gamma^\lambda_{\nu\kappa},\\
        \partial_t(\sqrt{-g}B^i)&=-\partial_j\left[\sqrt{-g}(b^ju^i - b^i u^j)\right],\\
        \frac{1}{\sqrt{-g}}\partial_i(\sqrt{-g}B^i)&=0,
    \end{aligned}
\end{equation}
where $\rho$ is the rest mass density, $u^\mu$ is the four-velocity, $\Gamma$ is the Christoffel symbol, $B^i$ and $b$ are the three- and four-magnetic fields, and $g$ is the metric determinant \citep[details see][]{2003ApJ...589..444G}. To avoid numerical problem at the BH horizon, Kerr-Schild coordinates are commonly used in GRMHD simulations \citep[e.g., ][]{Porth2017, 2022ApJS..259...64W}. In the simulations, an exponential radial coordinate system is used to improve the resolution near the BH horizon \citep{2003ApJ...589..444G}.

We initialize the simulation with a Fishbone-Moncrief hydrostatic equilibrium torus \citep{1976ApJ...207..962F}, embedding a single poloidal magnetic loop as a seed magnetic field.
Note that the torus remains in equilibrium only when aligned with the BH spin direction. To minimize the effects of this misalignment, following \cite{Liska2018, 2019ApJ...878...51W}, we tilt the disks by a relatively small angle ($25^\circ$) along the $y$-axis. The disk is also set to be relatively large, with an outer radius of $r_{\rm out} \sim 200\,r_{\rm g}$ (where $r_{\rm g} = GM/c^2$ is the gravitational radius and $M$ is the BH mass). While the accretion induced by deviations from equilibrium is present, it remains small and significantly weaker than the effects driven by the magnetic field. As the accretion rate of the HD model is a magnitude lower than the MHD models (see Fig~\ref{fig: Mdot}(a)).
The Kerr parameter $a$ is set to 0, and 0.9375 across all simulations to compare the effect of BH rotation.  
An ideal gas equation of state characterized by a constant adiabatic index of $\Gamma_{\rm g} = 5/3$ is utilized. 
We tried two different resolutions for the highly magnetized model {\tt T25a094nMAD}, with the higher grid resolution labeled with an extra {\tt H}.
\begin{table}[h]
\caption{The parameters of the GR(M)HD simulations.}
\centering
\begin{tabular}{lllllllll}
\hline
Models                                & a       & $\mathcal{T}_0\,(^\circ)$ & ${\tt lin\_frac}$ & ${\tt smoothness}$ & $r_{\rm d} \,[{r_{\rm g}}]$ & $A_\phi$               & $\beta_{\rm min}$ & Resolution              \\ \hline
{\tt T25a094HD}   & 0.9375  & 25         & 0.8            & 0.03             & 500                     & -                      & -                 & $144\times72\times96$   \\
{\tt T25a094wSANE}& 0.9375  & 25         & 0.8            & 0.03             & 1000                     & $A_{\phi}^{\tt wSANE}$ & 100               & $288\times128\times128$ \\
{\tt T25a094sSANE}& 0.9375  & 25         & 0.8            & 0.03             & 1000                     & $A_{\phi}^{\tt sSANE}$ & 100               & $288\times128\times128$ \\
{\tt T25a094nMAD} & 0.9375  & 25         & 0.9            & 0.02             & 2500                     & $A_{\phi}^{\tt MAD}$  & 100               & $288\times128\times128$ \\
{\tt T25a094nMADH} & 0.9375 & 25         & 0.9            & 0.02             & 2500                     & $A_{\phi}^{\tt MAD}$  & 100               & $392\times256\times192$ \\
{\tt T25a0nMAD}    & 0       & 25         & 0.8            & 0.03             & 1000                     & $A_{\phi}^{\tt MAD}$  & 100               & $288\times128\times128$ \\
 \hline
\end{tabular}\label{table:models}
\end{table}

We list the parameters of all simulations in this work in Table~\ref{table:models}. The naming convention for simulations is designed to encapsulate key parameters of the setup. Each model is labeled in the following format:
\begin{equation*}
\texttt{T[Tilt Angle]a[Spin] [Magnetic Configuration]}
\end{equation*}

\begin{itemize}
    \item \textbf{Tilt angle:} The initial tilt angle is represented by \texttt{T}, followed by its value in degrees. \texttt{T25} corresponds to a $25^\circ$ tilt angle.
    \item \textbf{BH spin:} The spin parameter $a$ is represented as \texttt{a[Value]}, where the value is the dimensionless spin parameter. Positive spin is written as, for example, \texttt{a094} ($a = 0.9375$), while negative spin is denoted as, for example, \texttt{a-094} ($a = -0.9375$).
    \item \textbf{Magnetic configuration:} The magnetic field configuration is specified at the end of the label, such as \texttt{nMAD} for a normal field strength of MAD setup or \texttt{sSANE} for a strong field strength of SANE setup, and \texttt{wSANE} for a weak field strength of SANE setup. 
\end{itemize}

For instance, the label \texttt{T25a094nMAD} corresponds to a torus with an initial tilt angle of $25^\circ$ (\texttt{T25}), a BH spin of $a = 0.94$ (\texttt{a094}), and a normal strength magnetic field for MAD (\texttt{nMAD}).

The initial magnetic field within the torus is supplied through the vector potential. To generate a purely poloidal magnetic field, we exclusively specify the toroidal component of the vector potential while keeping the poloidal component zero. Three different magnetic configurations ($A_\phi$ setup) are implemented, namely weak SANE ({\tt wSANE}), strong SANE ({\tt sSANE}), and MAD ({\tt nMAD}). The expressions of $A_\phi$ for them are written as follows:
\begin{equation}
\begin{aligned}
    A_{\phi}^{\rm \tt wSANE} &\propto {\rm max}[\rho-0.2,0],\\
    A_{\phi}^{\rm \tt sSANE} &\propto {\rm max}[(\rho-0.5)^3r^3,0],\\
    A_{\phi}^{\rm \tt MAD} &\propto {\rm max}\left[(r/r_{\rm in})^3\exp{(-r/400)}\rho-0.2,0\right].
\end{aligned}
\label{Eq: Aphi}
\end{equation}
The strength of the magnetic field is determined by the minimum plasma $\beta_{\rm min}$ in the torus, where plasma $\beta$ is the ratio between gas pressure ($p_{\rm gas}$) and magnetic pressure ($p_{\rm mag}$). To stimulate magnetorotational instability (MRI), a random perturbation to the initial torus is added to the internal energy $u\rightarrow u+\delta u$, $|\delta u/u|\leq u_{\rm jitter}$ \citep{2022ApJS..259...64W}. In this work, $u_{\rm jitter}$ is set to 0.1. 

\section{Numerical resolution and coordinate}\label{sec: resolution}

\begin{figure*}
    \centering
    \includegraphics[width=\linewidth]{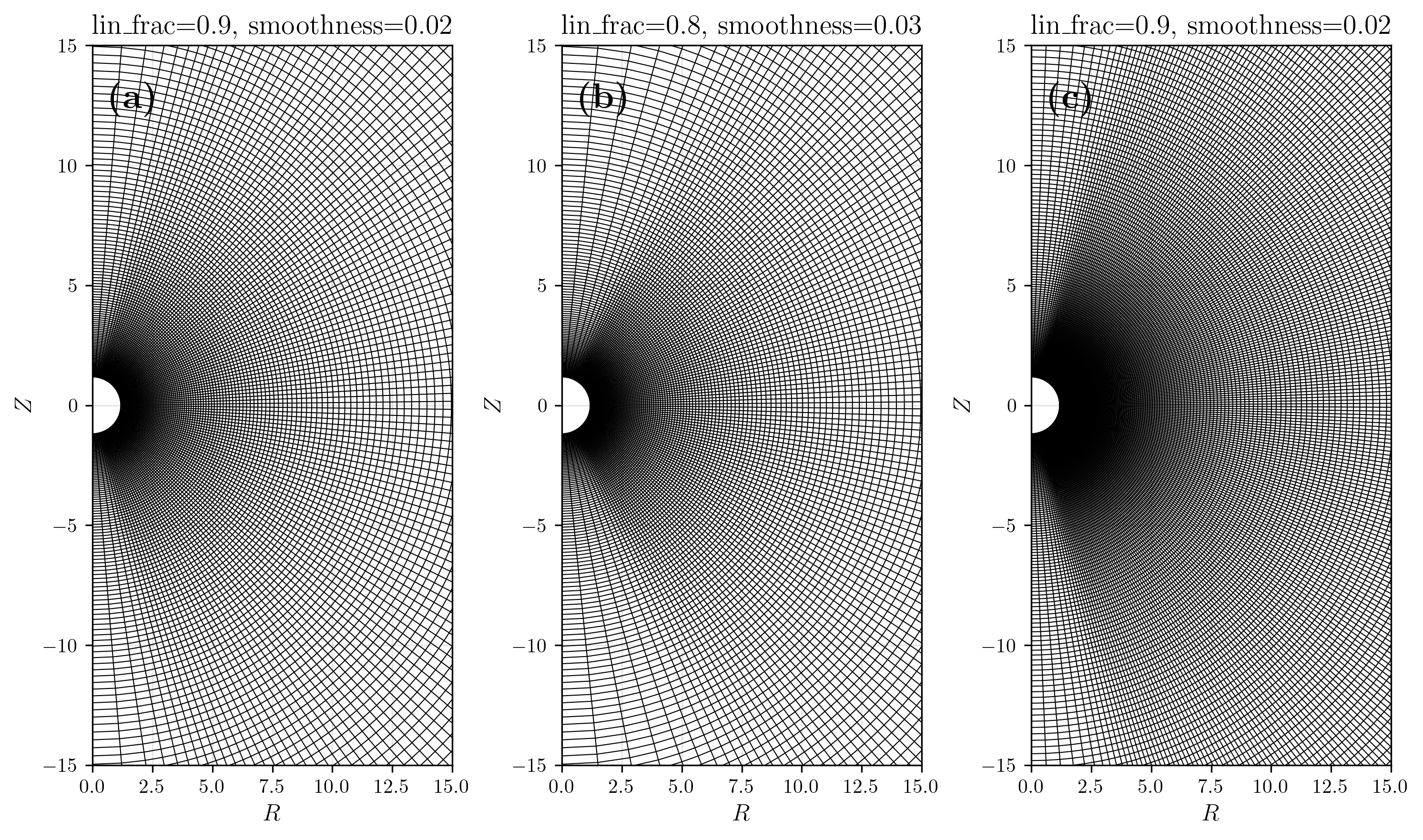}
    \caption{Panels (a) and (c) display the mesh grid configuration with parameters \texttt{lin\_frac} = 0.9 and \texttt{smoothness} = 0.02, corresponding to models \texttt{T25a094nMAD} and \texttt{T25a094nMADH}, respectively. In contrast, panel (b) shows the mesh grid used for the other MHD models, which employs \texttt{lin\_frac} = 0.8 and \texttt{smoothness} = 0.03.}
    \label{fig: mesh}
\end{figure*}

\begin{figure*}
    \centering
    \includegraphics[width=\linewidth]{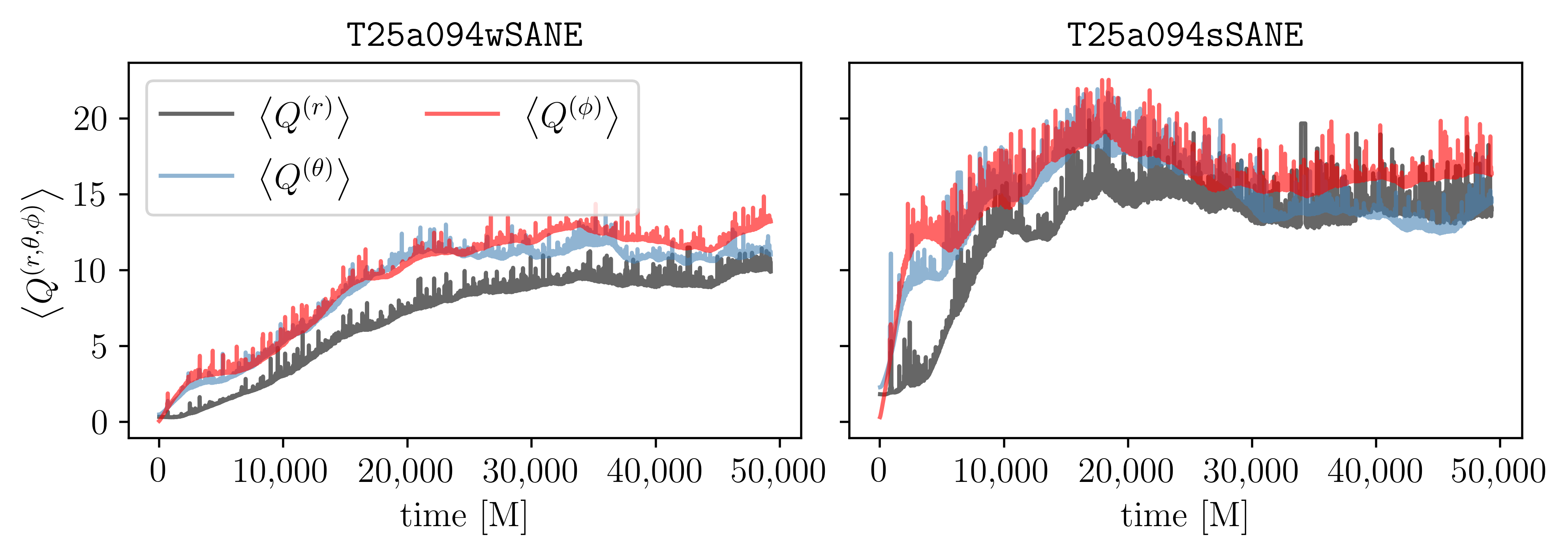}
    \caption{Time evolution of density averaged MRI quality factors $\langle Q^{(r,\theta,\phi)} \rangle$ for models {\tt T25a094wSANE} and {\tt T25a094sSANE} are shown in panels (a) and (b), respectively. The black, blue, and red lines represent $\langle Q^{(r)}\rangle$, $\langle Q^{(\theta)} \rangle$, and $\langle Q^{(\phi)} \rangle$, respectively.}    \label{fig: Q_factor}
\end{figure*}

In this study, we employ uniform wide-pole Kerr-Schild coordinates, as detailed in \cite{2024ApJ...977..200C}. This coordinate maintains relatively high resolution in most of the disk regions while allowing a sparser grid distribution in the polar regions. It increases the time step during simulations by a factor of 2. Considering the relatively small initial tilt used in our simulations, the low resolution in the polar region has no strong impact on the dynamics of precession studied in this work. In Fig.~\ref{fig: mesh}, we present the grid configurations with different parameter sets for the wide-pole Kerr-Schild coordinate.

We employ varying grid resolutions across different models, with a standard resolution of $288 \times 128 \times 128$ following \cite{2025ApJS..277...16D, event_horizon_telescope_collaboration_first_2022}. This resolution is sufficient to resolve the MRI, as demonstrated in Fig.~\ref{fig: Q_factor}. For the GRHD model {\tt T25a094HD}, the LT precession does not depend on the magnetic field and turbulence. Therefore, a reduced resolution of $144 \times 72 \times 96$ is used to save computational costs.

Following the simulation setup of {\tt KHARMA} in \cite{event_horizon_telescope_collaboration_first_2022}, we set the outer boundary at $r_{\rm d} = 1,000\,r_{\rm g}$ for most GRMHD models (see Table~\ref{table:models} for details). However, for the MAD model that generate powerful jets, we extend the simulation domain to $r_{\rm d} = 2,500\,r_{\rm g}$. Conversely, the GRHD model, which exhibits minimal outflows, uses a smaller domain with $r_{\rm d} = 500\,r_{\rm g}$ to maintain higher resolution. The polar boundary is set to be a transmitting boundary condition, which allows the plasma to pass through the polar boundary to reduce dissipation.

We present the density-weighted averaged MRI quality factors $\langle Q^{(r,\theta,\phi)}\rangle$ in $r$, $\theta$, and $\phi$ directions of two SANE models, {\tt T25a094wSANE} and {\tt T25a094sSANE} in Fig.~\ref{fig: Q_factor}. The calculation of $Q$ factors follows \cite{2019ApJS..243...26P}, and average is done within a range of $r<150\,r_{\rm g}$.
The averaged $Q^{(r)}$ and $Q^{(\theta)}$ reach the required value for resolving MRI suggested in \cite{Sorathia2012} ($Q^{(z)}\leq 10-15$, $Q^{(\phi)}\approx 10$). For the weakly magnetized model {\tt T25a094wSANE}, we have similar MRI qualification factors with \cite{2025ApJS..277...16D}. 

The magnetically driven precession mechanism shown in this work primarily relies on the large-scale poloidal magnetic field, rather than on a turbulent magnetic field produced by MRI, which requires rather high resolution to fully develop. Consequently, although the resolution is lower than that used in \cite{Liska2018}, our conclusions remain unaffected by the limited resolution.

\section{Measurement of disk and jet tilt and precession angles} \label{sec: measurement}

The measurement of tilt and precession angle of the disk and jet in this work follows the way used in previous studies \citep{2005ApJ...623..347F, Liska2018, 2019ApJ...878...51W}. Here we briefly introduce it.

The angular momentum vector of the disk can be written as \citep{2005ApJ...623..347F}
\begin{equation}
    (J_{\rm disk})_{\rho} = \frac{\epsilon_{\mu\nu\sigma\rho}L^{\mu\nu}S^{\sigma}}{2\sqrt{-S^\alpha S_\alpha}},
\end{equation}
where the total angular momentum is
\begin{equation}
    L^{\mu\nu} = \int\left(x^\mu T^{\nu 0} - x^\nu T^{\mu 0}\right) d^3x,
\end{equation}
with the four-momentum $S^\sigma$ written as
\begin{equation}
    S^\sigma=\int T^{\sigma0}d^3x.
\end{equation}
The fluid component of the energy-momentum tensor is $T^{\mu\nu}=\rho h u^\mu u^\nu + p g^{\mu\nu}$, where $h$ and $p$ ym{are} the enthalpy and gas pressure. 
Following \cite{Liska2018}, when calculating the disk angular momentum in Cartesian coordinates, the tilt angle of the disk $\mathcal{T}_{\rm disk}$ is obtained by  
\begin{equation}
    \mathcal{T}_{\rm disk}=\cos^{-1}\left(\frac{J_{\rm disk}^z}{|\mathbf{J_{\rm disk}}|}\right), \label{Eq: tilt}
\end{equation}
and precession angle $\mathcal{P}_{\rm disk}$ is
\begin{equation}
    \mathcal{P}_{\rm disk} = \tan^{-1}\left(J_{\rm disk}^y, J_{\rm disk}^x\right). \label{Eq: precession}
\end{equation}

The definition of jet orientation follows \cite{Liska2018}. We isolate the jet region using the criteria $r p_{\rm mag}/\rho > 0.5$, where $p_{\rm mag}$ is the magnetic pressure. Choosing the upper jet, for example, we measure the center of the upper jet in Cartesian coordinates by 
\begin{equation}
    x^i_{\rm jet}=\frac{\int p_{\rm mag}^{\rm up} x^i d^3x}{\int p_{\rm mag}^{\rm up}d^3x},
\end{equation}
where the $p_{\rm mag}^{\rm up}$ represent the magnetic pressure in the upper jet, and $i=1,2,3$. Then, the position angle of the jet can be obtained similarly with the disk in Eq.~\ref{Eq: tilt} and \ref{Eq: precession}.

\section{Calculations of torques: a toy model} \label{sec: Torque}

Here we review the calculation of the magnetic torque on the disk following \cite{Lai2003}. In a non-tilted disk threaded with the poloidal magnetic field, adopting the symbols in \cite{Lai2003}, the surface current in the disk is expressed as
\begin{equation}
    K_{\phi}=\int J_{\phi} dz = \frac{c}{2\pi}B_{\rm R}^+, \label{Eq: Kphi} 
\end{equation}
where $J_\phi$ is the current density, $B_{\rm R}^+\equiv B_{Z} \tan{\theta}$ is the magnetic field on the upper disk surface, and pitch angle of the poloidal magnetic field $\theta=\tan^{-1}|B_{\rm R}^+/B_{Z}|$.
For a non-tilted disk, the magnetic force is given by  $F_{\rm mag} = K_\phi \times B_{Z}$, which is purely radial and does not produce any torque for precession.

In the tilted disk, assuming that the rotating BH has aligned the magnetic field with its spin direction while the disk remains misaligned. The tilt angle for the disk is $\tilde{\beta}$. We define a coordinate system with the $z$-axis aligned with the disk angular momentum. The unit vector in this direction is denoted as ${\hat{l}}$, and ${\hat{Z}}$ is the unit vector of the $Z$-axis in the non-tilted coordinate.
Now the vertical magnetic field $\mathbf{B_{Z}}$ has two components in the tilted coordinate:
\begin{equation}
    \mathbf{B_{Z}} = B_Z \cos{\tilde{\beta}} {\hat{l}} + B_Z \sin{\tilde{\beta}}\sin{\phi} {\hat{r}}.
    \label{Eq: Bz_decompose}
\end{equation}
Hence, the magnetic force that is vertical to the disk is given by
\begin{equation}
    F_z=-\frac{1}{c}K_\phi B_Z \sin{\tilde{\beta}}\sin{\phi}.
\end{equation}
For a small tilt, Equation~\ref{Eq: Kphi} remains approximately valid.
Integrate over $\phi$ direction, the torque per unit area is given by
\begin{equation}
    \left<\mathbf{T_{\rm prec}}\right>=-\frac{1}{2c}r K_\phi B_Z {\hat{Z}}\times{\hat{l}} 
    =-\frac{1}{4\pi}rB_Z^2\tan{\theta}{\hat{Z}}\times{\hat{l}}.
\end{equation}
The total torque on the disk $\mathbf{T_{\rm tot}}$ is 
\begin{equation}
    \mathbf{T_{\rm tot}}=\int_{r_{\rm in}}^{r_{\rm out}}2\pi rdr\left<\mathbf{T_{\rm prec}}\right>=-\int_{r_{\rm in}}^{r_{\rm out}}dr \frac{1}{2}r^2 B_Z^2\tan{\theta} {\hat{Z}}\times{\hat{l}}. \label{Eq: mag_torq}
\end{equation}
Thus, the angular frequency of the magnetically driven precession is
\begin{equation}
    \mathbf{\Omega_{\rm prec}} = -\frac{1}{L_{\rm disk}}\int_{r_{\rm in}}^{r_{\rm out}}dr \frac{1}{2}r^2B_Z^2\tan{\theta} {\hat{Z}},
\end{equation}
where $L_{\rm disk} = \int_{r_{\rm in}}^{r_{\rm out}}\Sigma r^2 \Omega_{\rm d} 2\pi r dr$, is the disk angular momentum, $\Sigma$ is the surface density, and $\Omega_{\rm d}$ is the disk angular frequency.

In realistic situation, such strong magnetic alignment may not occur. However, due to the influence of the rotating BH, the orientation of the poloidal magnetic field, including the jet direction, typically exhibits some deviation from the disk orientation, which satisfies the requirement of this simple model. 

As many works have indicated, in the tilted disk situation, the disk may undergo LT precession, which is prograde \citep[e.g.,][]{2024Natur.630..325P}. From the previous discussion, we see that in weakly magnetized disks, magnetic torque slows down the LT precession. The LT torque per unit area is given as \citep{McKinney2013}
\begin{equation}
T_{\rm LT} = \sin\tilde{\beta} \Omega_{\rm LT} L_{\rm disk}, 
\end{equation}
with
\begin{equation}
\Omega_{\rm LT}=\frac{1}{L_{\rm disk}} \int_{r_{\rm in}}^{r_{\rm out}}\frac{2M^2 a}{r^3}r^2\Omega_{\rm d}\Sigma 2\pi r dr.
\end{equation}

\section{Disk size, accretion rate and magnetic flux} \label{sec: Mdot}

\begin{figure}
    \centering
    \includegraphics[width=\linewidth]{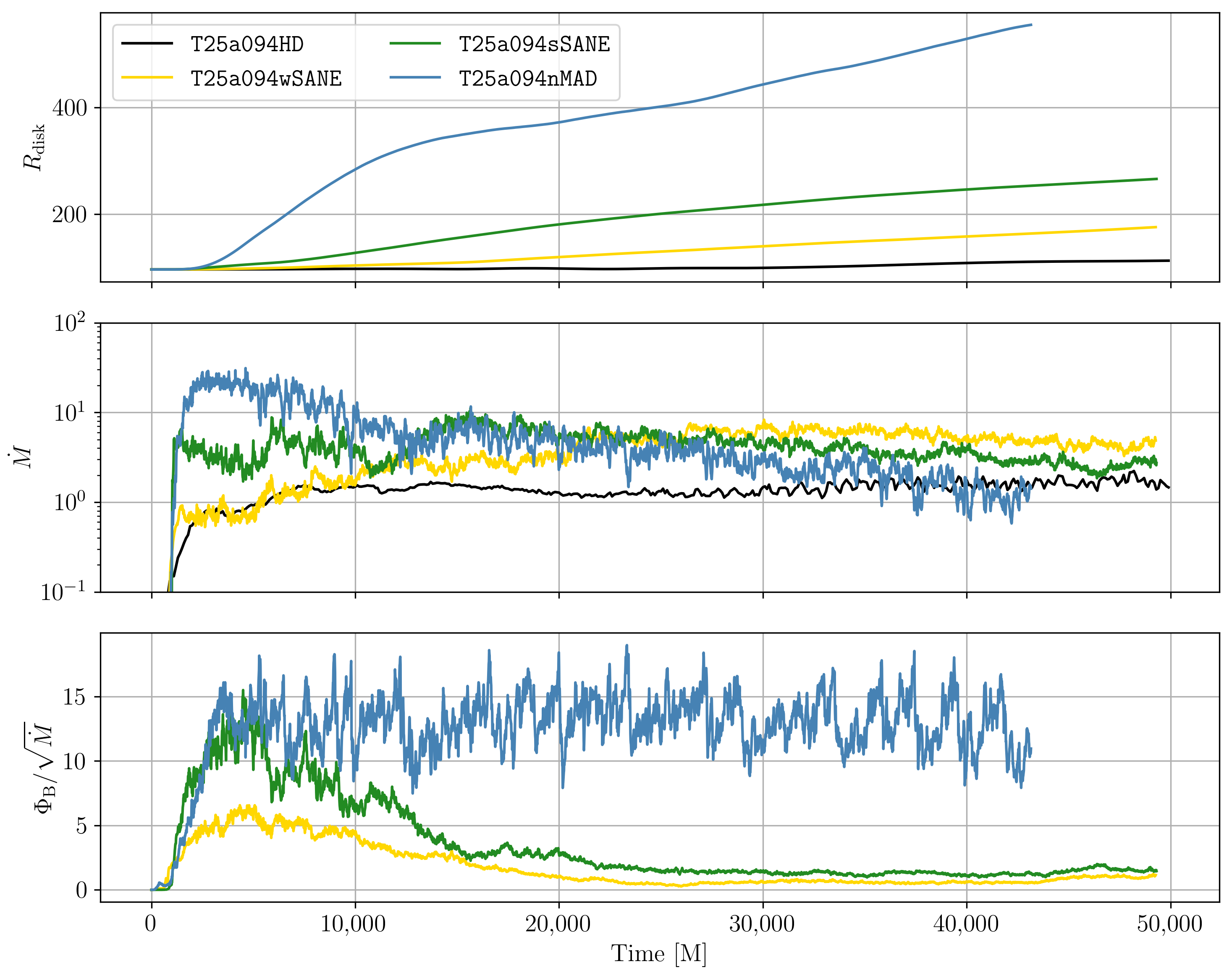}
    \caption{Panels (a), (b), and (c) show the time evolutions of disk size, accretion rate, and scaled magnetic flux rate, respectively, for different models: {\tt T25a094HD} (black), {\tt T25a094wSANE} (yellow), {\tt T25a094sSANE} (green), and {\tt T25a094nMAD} (blue).}    \label{fig: Mdot}
\end{figure}

In Fig.~\ref{fig: Mdot}, we present the averaged disk radius $R_{\rm disk}$, accretion rate $\dot M$, and scaled magnetic flux rate $\Phi_{\rm B}/\sqrt{\dot M}$ on the event horizon in panels (a)-(c), respectively. The definition of the $R_{\rm disk}$ follows \cite{Liska2018}, and the $\dot M$ and $\Phi_{\rm B}$ follows \cite{2019ApJS..243...26P}, which are given by
\begin{equation}
    \begin{aligned}
        R_{\rm disk} &= \frac{\int\int_0^{\pi}\int_0^{2\pi} r \rho\sqrt{-g}drd\theta d\phi}{\int\int_0^{\pi}\int_0^{2\pi} \rho\sqrt{-g}drd\theta d\phi},\\
        \dot{M} &= \int_0^{2\pi} \int_0^{\pi} \rho u^r \sqrt{-g} \, d\theta \, d\phi,\\
        \Phi_{\rm B} &= \frac{1}{2} \int_0^{2\pi} \int_{0}^{\pi} \left|B^r\right| \sqrt{-g} \, d\theta \, d\phi,
    \end{aligned}
\end{equation}
where in the calculation of $R_{\rm disk}$, we only pick the region with $\rho>10^{-5}$, and $g$ is the metric determinant.

From Fig.~\ref{fig: Mdot}(a), the disk size grows significantly faster in the strongly magnetized model {\tt T25a094nMAD} compared to SANE models, while it remains nearly unchanged in the non-magnetized model {\tt T25a094HD}. This is due to the much stronger outflows in the strongly magnetized model {\tt T25a094nMAD}. In contrast, the torus in the non-magnetized model {\tt T25a094HD} maintains most of the plasma in the torus near its initial position.

Since the smooth transition between the torus and atmosphere, minor accretion still occurs in non-magnetized model {\tt T25a094HD} from the edge of the torus, as shown in Fig.~\ref{fig: Mdot}(b). However, their accretion rates remain significantly lower than those in most of the MHD models. This ensures that our results are not affected by this minor accretion. Notably, the torus of model {\tt T25a094nMAD} undergo significant expansion during the simulation, causing the disk density to decrease by more than an order of magnitude. As a result, by the end of the simulation, their accretion rates fall below those of the non-magnetized model {\tt T25a094HD}.

Both SANE models exhibit a low dimensionless magnetic flux, remaining below 1 in code units. Higlly magnetized model {\tt T25a094nMAD} meets the standard MAD criterion with $\Phi_{\rm B}/\sqrt{\dot{M}} \sim 15$ \citep{2022MNRAS.511.2040B, 2011MNRAS.418L..79T}.

\section{Comparison between high and low resolutions}
\label{sec: res_comp}

\begin{figure}
    \centering
    \includegraphics[width=\linewidth]{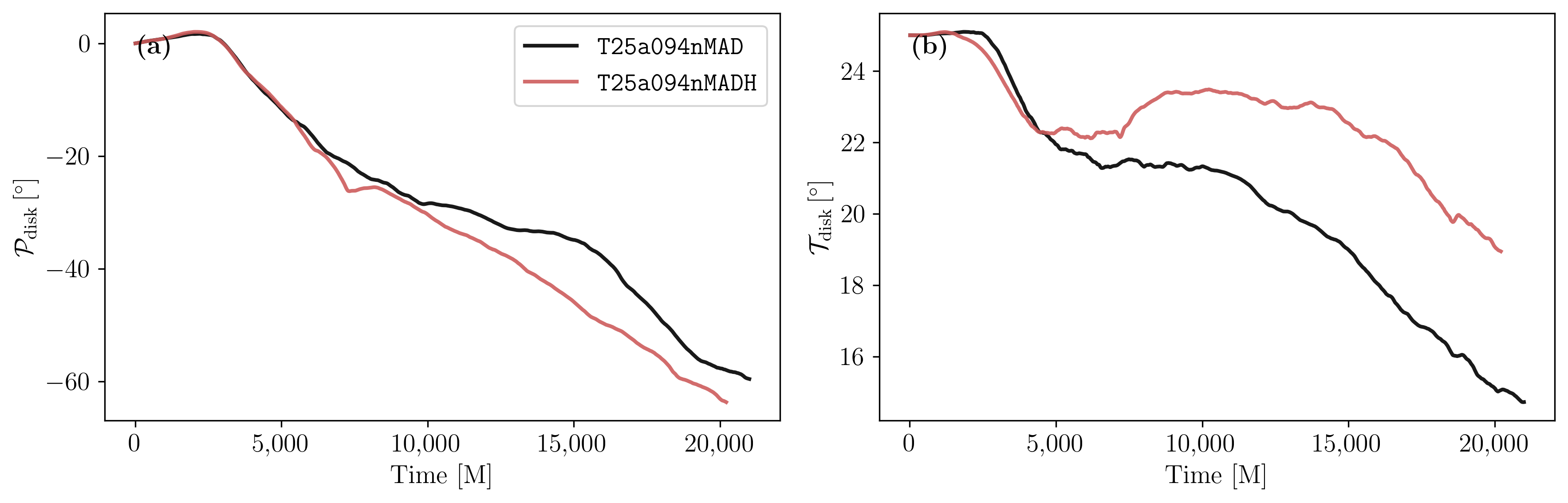}
    \caption{Comparison of the disk precession angle \(\mathcal{P}_{\text{disk}}\) and tilt angle \(\mathcal{T}_{\text{disk}}\) from simulations {\tt T25a094nMAD} (standard resolution) and {\tt T25a094nMADH} (higher resolution).}    \label{fig: res_comp}
\end{figure}

In Fig.~\ref{fig: res_comp}, we perform a resolution convergence test comparing the standard-resolution model {\tt T25a094nMAD} with its high-resolution counterpart {\tt T25a094nMADH}. The time evolution of precession angle ($\mathcal{P}_{\mathrm{disk}}$) shows excellent agreement between both models, while the time evolution of tilt angle ($\mathcal{T}_{\mathrm{disk}}$) exhibits similar behavior albeit with slightly slower alignment in the high-resolution model. These results demonstrate both the robustness of retrograde behavior in MAD flows and the adequacy of our standard resolution for capturing the system's precession dynamics.


\bibliography{sample701}{}

\begin{thebibliography}{}
\expandafter\ifx\csname natexlab\endcsname\relax\def\natexlab#1{#1}\fi
\providecommand{\url}[1]{\href{#1}{#1}}
\providecommand{\dodoi}[1]{doi:~\href{http://doi.org/#1}{\nolinkurl{#1}}}
\providecommand{\doeprint}[1]{\href{http://ascl.net/#1}{\nolinkurl{http://ascl.net/#1}}}
\providecommand{\doarXiv}[1]{\href{https://arxiv.org/abs/#1}{\nolinkurl{https://arxiv.org/abs/#1}}}

\bibitem[{K. Akiyama {et~al.}(2019)Akiyama, Alberdi, Alef, Asada, Azulay, Baczko, Ball, Balokovi{\'{c}}, Barrett, Bintley, Blackburn, Boland, Bouman, Bower, Bremer, Brinkerink, Brissenden, Britzen, Broderick, Broguiere, Bronzwaer, Byun, Carlstrom, Chael, Chan, Chatterjee, Chatterjee, Chen, {Chen 陈}, Cho, Christian, Conway, Cordes, Crew, Cui, Davelaar, {De Laurentis}, Deane, Dempsey, Desvignes, Dexter, Doeleman, Eatough, Falcke, Fish, Fomalont, Fraga-Encinas, Friberg, Fromm, G{\'{o}}mez, Galison, Gammie, Garc{\'{i}}a, Gentaz, Georgiev, Goddi, Gold, {Gu 顾}, Gurwell, Hada, Hecht, Hesper, {Ho 何}, Ho, Honma, Huang, {Huang 黄}, Hughes, Ikeda, Inoue, Issaoun, James, Jannuzi, Janssen, Jeter, {Jiang 江}, Johnson, Jorstad, Jung, Karami, Karuppusamy, Kawashima, Keating, Kettenis, Kim, Kim, Kim, Kino, Koay, Koch, Koyama, Kramer, Kramer, Krichbaum, Kuo, Lauer, Lee, {Li 李}, {Li 李}, Lindqvist, Liu, Liuzzo, Lo, Lobanov, Loinard, Lonsdale, {Lu 路}, MacDonald, {Mao 毛}, Markoff, Marrone, Marscher,
  Mart{\'{i}}-Vidal, Matsushita, Matthews, Medeiros, Menten, Mizuno, Mizuno, Moran, Moriyama, Moscibrodzka, Mul̈ler, Nagai, Nagar, Nakamura, Narayan, Narayanan, Natarajan, Neri, Ni, Noutsos, Okino, Olivares, Oyama, {\"{O}}zel, Palumbo, Patel, Pen, Pesce, Pi{\'{e}}tu, Plambeck, PopStefanija, Porth, Prather, Preciado-L{\'{o}}pez, Psaltis, Pu, Ramakrishnan, Rao, Rawlings, Raymond, Rezzolla, Ripperda, Roelofs, Rogers, Ros, Rose, Roshanineshat, Rottmann, Roy, Ruszczyk, Ryan, Rygl, S{\'{a}}nchez, S{\'{a}}nchez-Arguelles, Sasada, Savolainen, Schloerb, Schuster, Shao, {Shen 沈}, Small, Sohn, SooHoo, Tazaki, Tiede, Tilanus, Titus, Toma, Torne, Trent, Trippe, Tsuda, van Bemmel, van Langevelde, van Rossum, Wagner, Wardle, Weintroub, Wex, Wharton, Wielgus, Wong, {Wu 吴}, Young, Young, Younsi, {Yuan 袁}, {Yuan 袁}, Zensus, Zhao, Zhao, Zhu, Anczarski, Baganoff, Eckart, Farah, Haggard, Meyer-Zhao, Michalik, Nadolski, Neilsen, Nishioka, Nowak, Pradel, Primiani, Souccar, Vertatschitsch, Yamaguchi, \& Zhang}]{Akiyama2019}
Akiyama, K., Alberdi, A., Alef, W., {et~al.} 2019, \bibinfo{title}{{First M87 Event Horizon Telescope Results. V. Physical Origin of the Asymmetric Ring},} The Astrophysical Journal Letters, 875, L5, \dodoi{10.3847/2041-8213/ab0f43}

\bibitem[{Z.~L. Andalman {et~al.}(2022)Andalman, Liska, Tchekhovskoy, Coughlin, \& Stone}]{Andalman2022}
Andalman, Z.~L., Liska, M.~T., Tchekhovskoy, A., Coughlin, E.~R., \& Stone, N. 2022, \bibinfo{title}{{Tidal disruption discs formed and fed by stream-stream and stream-disc interactions in global GRHD simulations},} Monthly Notices of the Royal Astronomical Society, 510, 1627, \dodoi{10.1093/mnras/stab3444}

\bibitem[{J.~M. {Bardeen} \& J.~A. {Petterson}(1975){Bardeen} \& {Petterson}}]{1975ApJ...195L..65B}
{Bardeen}, J.~M., \& {Petterson}, J.~A. 1975, \bibinfo{title}{{The Lense-Thirring Effect and Accretion Disks around Kerr Black Holes},} The Astrophysical Journall, 195, L65, \dodoi{10.1086/181711}

\bibitem[{S.~S. {Bavera} {et~al.}(2020){Bavera}, {Fragos}, {Qin}, {Zapartas}, {Neijssel}, {Mandel}, {Batta}, {Gaebel}, {Kimball}, \& {Stevenson}}]{2020A&A...635A..97B}
{Bavera}, S.~S., {Fragos}, T., {Qin}, Y., {et~al.} 2020, \bibinfo{title}{{The origin of spin in binary black holes. Predicting the distributions of the main observables of Advanced LIGO},} \aap, 635, A97, \dodoi{10.1051/0004-6361/201936204}

\bibitem[{M.~C. {Begelman} {et~al.}(2022){Begelman}, {Scepi}, \& {Dexter}}]{2022MNRAS.511.2040B}
{Begelman}, M.~C., {Scepi}, N., \& {Dexter}, J. 2022, \bibinfo{title}{{What really makes an accretion disc MAD},} \mnras, 511, 2040, \dodoi{10.1093/mnras/stab3790}

\bibitem[{A. {Caproni} {et~al.}(2007){Caproni}, {Abraham}, {Livio}, \& {Mosquera Cuesta}}]{2007MNRAS.379..135C}
{Caproni}, A., {Abraham}, Z., {Livio}, M., \& {Mosquera Cuesta}, H.~J. 2007, \bibinfo{title}{{Is the Bardeen-Petterson effect responsible for the warping and precession in NGC4258?},} Monthly Notices of the Royal Astronomical Society, 379, 135, \dodoi{10.1111/j.1365-2966.2007.11918.x}

\bibitem[{A. {Caproni} {et~al.}(2006){Caproni}, {Abraham}, \& {Mosquera Cuesta}}]{2006ApJ...638..120C}
{Caproni}, A., {Abraham}, Z., \& {Mosquera Cuesta}, H.~J. 2006, \bibinfo{title}{{Bardeen-Petterson Effect and the Disk Structure of the Seyfert Galaxy NGC 1068},} The Astrophysical Journal, 638, 120, \dodoi{10.1086/498684}

\bibitem[{K. Chatterjee {et~al.}(2023)Chatterjee, Liska, Tchekhovskoy, \& Markoff}]{Chatterjee2023}
Chatterjee, K., Liska, M., Tchekhovskoy, A., \& Markoff, S. 2023, \bibinfo{title}{{Misaligned magnetized accretion flows onto spinning black holes: magneto-spin alignment, outflow power and intermittent jets},} \doarXiv{2311.00432}

\bibitem[{H. {Cho} {et~al.}(2024){Cho}, {Prather}, {Su}, {Narayan}, \& {Natarajan}}]{2024ApJ...977..200C}
{Cho}, H., {Prather}, B.~S., {Su}, K.-Y., {Narayan}, R., \& {Natarajan}, P. 2024, \bibinfo{title}{{Multizone Modeling of Black Hole Accretion and Feedback in 3D GRMHD: Bridging Vast Spatial and Temporal Scales},} \apj, 977, 200, \dodoi{10.3847/1538-4357/ad9561}

\bibitem[{Y. Cui {et~al.}(2023)Cui, Hada, Kawashima, Kino, Lin, Mizuno, Ro, Honma, Yi, Yu, Park, Jiang, Shen, Kravchenko, Algaba, Cheng, Cho, Giovannini, Giroletti, Jung, Lu, Niinuma, Oh, Ohsuga, Sawada-Satoh, Sohn, Takahashi, Takamura, Tazaki, Trippe, Wajima, Akiyama, An, Asada, Buttaccio, Byun, Cui, Hagiwara, Hirota, Hodgson, Kawaguchi, Kim, Lee, Lee, Lee, Maccaferri, Melis, Melnikov, Migoni, Oh, Sugiyama, Wang, Zhang, Chen, Hwang, Jung, Kim, Kim, Kobayashi, Li, Li, Li, Liu, Liu, Liu, Oh, Oyama, Roh, Wang, Wang, Wang, Xia, Yan, Yeom, Yonekura, Yuan, Zhang, Zhao, \& Zhong}]{Cui2023}
Cui, Y., Hada, K., Kawashima, T., {et~al.} 2023, \bibinfo{title}{{Precessing jet nozzle connecting to a spinning black hole in M87},} Nature, 621, 711, \dodoi{10.1038/s41586-023-06479-6}

\bibitem[{V. {Dhruv} {et~al.}(2025){Dhruv}, {Prather}, {Wong}, \& {Gammie}}]{2025ApJS..277...16D}
{Dhruv}, V., {Prather}, B., {Wong}, G.~N., \& {Gammie}, C.~F. 2025, \bibinfo{title}{{A Survey of General Relativistic Magnetohydrodynamic Models for Black Hole Accretion Systems},} \apjs, 277, 16, \dodoi{10.3847/1538-4365/adaea6}

\bibitem[{M. {Elvis} {et~al.}(2002){Elvis}, {Risaliti}, \& {Zamorani}}]{2002ApJ...565L..75E}
{Elvis}, M., {Risaliti}, G., \& {Zamorani}, G. 2002, \bibinfo{title}{{Most Supermassive Black Holes Must Be Rapidly Rotating},} \apjl, 565, L75, \dodoi{10.1086/339197}

\bibitem[{ {Event Horizon Telescope Collaboration} {et~al.}(2022){Event Horizon Telescope Collaboration}, Akiyama, Alberdi, Alef, Algaba, Anantua, Asada, Azulay, Bach, Baczko, Ball, Baloković, Barrett, Bauböck, Benson, Bintley, Blackburn, Blundell, Bouman, Bower, Boyce, Bremer, Brinkerink, Brissenden, Britzen, Broderick, Broguiere, Bronzwaer, Bustamante, Byun, Carlstrom, Ceccobello, Chael, Chan, Chatterjee, Chatterjee, Chen, Chen, Cheng, Cho, Christian, Conroy, Conway, Cordes, Crawford, Crew, Cruz-Osorio, Cui, Davelaar, De~Laurentis, Deane, Dempsey, Desvignes, Dexter, Dhruv, Doeleman, Dougal, Dzib, Eatough, Emami, Falcke, Farah, Fish, Fomalont, Ford, Fraga-Encinas, Freeman, Friberg, Fromm, Fuentes, Galison, Gammie, García, Gentaz, Georgiev, Goddi, Gold, Gómez-Ruiz, Gómez, Gu, Gurwell, Hada, Haggard, Haworth, Hecht, Hesper, Heumann, Ho, Ho, Honma, Huang, Huang, Hughes, Ikeda, Violette~Impellizzeri, Inoue, Issaoun, James, Jannuzi, Janssen, Jeter, Jiang, Jiménez-Rosales, Johnson, Jorstad, Joshi, Jung,
  Karami, Karuppusamy, Kawashima, Keating, Kettenis, Kim, Kim, Kim, Kim, Kino, Koay, Kocherlakota, Kofuji, Koch, Koyama, Kramer, Kramer, Krichbaum, Kuo, Bella, Lauer, Lee, Lee, Leung, Levis, Li, Lico, Lindahl, Lindqvist, Lisakov, Liu, Liu, Liuzzo, Lo, Lobanov, Loinard, Lonsdale, Lu, Mao, Marchili, Markoff, Marrone, Marscher, Martí-Vidal, Matsushita, Matthews, Medeiros, Menten, Michalik, Mizuno, Mizuno, Moran, Moriyama, Moscibrodzka, Müller, Mus, Musoke, Myserlis, Nadolski, Nagai, Nagar, Nakamura, Narayan, Narayanan, Natarajan, Nathanail, Navarro~Fuentes, Neilsen, Neri, Ni, Noutsos, Nowak, Oh, Okino, Olivares, Ortiz-León, Oyama, Özel, Palumbo, Filippos~Paraschos, Park, Parsons, Patel, Pen, Pesce, Piétu, Plambeck, PopStefanija, Porth, Pötzl, Prather, Preciado-López, Psaltis, Pu, Ramakrishnan, Rao, Rawlings, Raymond, Rezzolla, Ricarte, Ripperda, Roelofs, Rogers, Ros, Romero-Cañizales, Roshanineshat, Rottmann, Roy, Ruiz, Ruszczyk, Rygl, Sánchez, Sánchez-Argüelles, Sánchez-Portal, Sasada, Satapathy,
  Savolainen, Schloerb, Schonfeld, Schuster, Shao, Shen, Small, Sohn, SooHoo, Souccar, Sun, Tazaki, Tetarenko, Tiede, Tilanus, Titus, Torne, Traianou, Trent, Trippe, Turk, van Bemmel, van Langevelde, van Rossum, Vos, Wagner, Ward-Thompson, Wardle, Weintroub, Wex, Wharton, Wielgus, Wiik, Witzel, Wondrak, Wong, Wu, Yamaguchi, Yoon, Young, Young, Younsi, Yuan, Yuan, Zensus, Zhang, Zhao, Zhao, Chan, Qiu, Ressler, \& White}]{event_horizon_telescope_collaboration_first_2022}
{Event Horizon Telescope Collaboration}, Akiyama, K., Alberdi, A., {et~al.} 2022, \bibinfo{title}{First {Sagittarius} {A}* {Event} {Horizon} {Telescope} {Results}. {V}. {Testing} {Astrophysical} {Models} of the {Galactic} {Center} {Black} {Hole},} The Astrophysical Journal, 930, L16, \dodoi{10.3847/2041-8213/ac6672}

\bibitem[{L.~G. {Fishbone} \& V. {Moncrief}(1976){Fishbone} \& {Moncrief}}]{1976ApJ...207..962F}
{Fishbone}, L.~G., \& {Moncrief}, V. 1976, \bibinfo{title}{{Relativistic fluid disks in orbit around Kerr black holes.},} The Astrophysical Journal, 207, 962, \dodoi{10.1086/154565}

\bibitem[{P.~C. {Fragile} \& P. {Anninos}(2005){Fragile} \& {Anninos}}]{2005ApJ...623..347F}
{Fragile}, P.~C., \& {Anninos}, P. 2005, \bibinfo{title}{{Hydrodynamic Simulations of Tilted Thick-Disk Accretion onto a Kerr Black Hole},} The Astrophysical Journal, 623, 347, \dodoi{10.1086/428433}

\bibitem[{P.~C. {Fragile} \& O.~M. {Blaes}(2008){Fragile} \& {Blaes}}]{2008ApJ...687..757F}
{Fragile}, P.~C., \& {Blaes}, O.~M. 2008, \bibinfo{title}{{Epicyclic Motions and Standing Shocks in Numerically Simulated Tilted Black Hole Accretion Disks},} The Astrophysical Journal, 687, 757, \dodoi{10.1086/591936}

\bibitem[{P.~C. {Fragile} {et~al.}(2007){Fragile}, {Blaes}, {Anninos}, \& {Salmonson}}]{2007ApJ...668..417F}
{Fragile}, P.~C., {Blaes}, O.~M., {Anninos}, P., \& {Salmonson}, J.~D. 2007, \bibinfo{title}{{Global General Relativistic Magnetohydrodynamic Simulation of a Tilted Black Hole Accretion Disk},} The Astrophysical Journal, 668, 417, \dodoi{10.1086/521092}

\bibitem[{P.~C. {Fragile} \& M. {Liska}(2024){Fragile} \& {Liska}}]{2024arXiv240410052F}
{Fragile}, P.~C., \& {Liska}, M. 2024, \bibinfo{title}{{Tilted Accretion Disks},} arXiv e-prints, arXiv:2404.10052, \dodoi{10.48550/arXiv.2404.10052}

\bibitem[{T. {Fragos} {et~al.}(2010){Fragos}, {Tremmel}, {Rantsiou}, \& {Belczynski}}]{2010ApJ...719L..79F}
{Fragos}, T., {Tremmel}, M., {Rantsiou}, E., \& {Belczynski}, K. 2010, \bibinfo{title}{{Black Hole Spin-Orbit Misalignment in Galactic X-ray Binaries},} \apjl, 719, L79, \dodoi{10.1088/2041-8205/719/1/L79}

\bibitem[{A. {Franchini} {et~al.}(2016){Franchini}, {Lodato}, \& {Facchini}}]{2016MNRAS.455.1946F}
{Franchini}, A., {Lodato}, G., \& {Facchini}, S. 2016, \bibinfo{title}{{Lense-Thirring precession around supermassive black holes during tidal disruption events},} \mnras, 455, 1946, \dodoi{10.1093/mnras/stv2417}

\bibitem[{C.~F. {Gammie} {et~al.}(2003){Gammie}, {McKinney}, \& {T{\'o}th}}]{2003ApJ...589..444G}
{Gammie}, C.~F., {McKinney}, J.~C., \& {T{\'o}th}, G. 2003, \bibinfo{title}{{HARM: A Numerical Scheme for General Relativistic Magnetohydrodynamics},} The Astrophysical Journal, 589, 444, \dodoi{10.1086/374594}

\bibitem[{L. {Gou} {et~al.}(2011){Gou}, {McClintock}, {Reid}, {Orosz}, {Steiner}, {Narayan}, {Xiang}, {Remillard}, {Arnaud}, \& {Davis}}]{2011ApJ...742...85G}
{Gou}, L., {McClintock}, J.~E., {Reid}, M.~J., {et~al.} 2011, \bibinfo{title}{{The Extreme Spin of the Black Hole in Cygnus X-1},} \apj, 742, 85, \dodoi{10.1088/0004-637X/742/2/85}

\bibitem[{P.~T. {Kondratko} {et~al.}(2005){Kondratko}, {Greenhill}, \& {Moran}}]{2005ApJ...618..618K}
{Kondratko}, P.~T., {Greenhill}, L.~J., \& {Moran}, J.~M. 2005, \bibinfo{title}{{Evidence for a Geometrically Thick Self-Gravitating Accretion Disk in NGC 3079},} The Astrophysical Journal, 618, 618, \dodoi{10.1086/426101}

\bibitem[{A. {Kotrlov{\'a}} {et~al.}(2020){Kotrlov{\'a}}, {{\v{S}}r{\'a}mkov{\'a}}, {T{\"o}r{\"o}k}, {Goluchov{\'a}}, {Hor{\'a}k}, {Straub}, {Lan{\v{c}}ov{\'a}}, {Stuchl{\'\i}k}, \& {Abramowicz}}]{2020A&A...643A..31K}
{Kotrlov{\'a}}, A., {{\v{S}}r{\'a}mkov{\'a}}, E., {T{\"o}r{\"o}k}, G., {et~al.} 2020, \bibinfo{title}{{Models of high-frequency quasi-periodic oscillations and black hole spin estimates in Galactic microquasars},} \aap, 643, A31, \dodoi{10.1051/0004-6361/201937097}

\bibitem[{D. {Lai}(1999){Lai}}]{1999ApJ...524.1030L}
{Lai}, D. 1999, \bibinfo{title}{{Magnetically Driven Warping, Precession, and Resonances in Accretion Disks},} \apj, 524, 1030, \dodoi{10.1086/307850}

\bibitem[{D. Lai(2003)Lai}]{Lai2003}
Lai, D. 2003, \bibinfo{title}{{Warping of Accretion Disks with Magnetically Driven Outflows: A Possible Origin for Jet Precession},} The Astrophysical Journal, 591, L119, \dodoi{10.1086/377163}

\bibitem[{M. Liska {et~al.}(2018)Liska, Hesp, Tchekhovskoy, Ingram, van~der Klis, \& Markoff}]{Liska2018}
Liska, M., Hesp, C., Tchekhovskoy, A., {et~al.} 2018, \bibinfo{title}{{Formation of precessing jets by tilted black hole discs in 3D general relativistic MHD simulations},} Monthly Notices of the Royal Astronomical Society: Letters, 474, L81, \dodoi{10.1093/mnrasl/slx174}

\bibitem[{M. {Liska} {et~al.}(2021){Liska}, {Hesp}, {Tchekhovskoy}, {Ingram}, {van der Klis}, {Markoff}, \& {Van Moer}}]{2021MNRAS.507..983L}
{Liska}, M., {Hesp}, C., {Tchekhovskoy}, A., {et~al.} 2021, \bibinfo{title}{{Disc tearing and Bardeen-Petterson alignment in GRMHD simulations of highly tilted thin accretion discs},} Monthly Notices of the Royal Astronomical Society, 507, 983, \dodoi{10.1093/mnras/staa099}

\bibitem[{M. {Liska} {et~al.}(2019){Liska}, {Tchekhovskoy}, {Ingram}, \& {van der Klis}}]{2019MNRAS.487..550L}
{Liska}, M., {Tchekhovskoy}, A., {Ingram}, A., \& {van der Klis}, M. 2019, \bibinfo{title}{{Bardeen-Petterson alignment, jets, and magnetic truncation in GRMHD simulations of tilted thin accretion discs},} Monthly Notices of the Royal Astronomical Society, 487, 550, \dodoi{10.1093/mnras/stz834}

\bibitem[{P.~L. {Luque-Escamilla} {et~al.}(2015){Luque-Escamilla}, {Mart{\'\i}}, \& {Mart{\'\i}nez-Aroza}}]{2015A&A...584A.122L}
{Luque-Escamilla}, P.~L., {Mart{\'\i}}, J., \& {Mart{\'\i}nez-Aroza}, J. 2015, \bibinfo{title}{{The precessing jets of <ASTROBJ>1E 1740.7-2942</ASTROBJ>},} \aap, 584, A122, \dodoi{10.1051/0004-6361/201527238}

\bibitem[{B. {Margon}(1984){Margon}}]{1984ARAA..22..507M}
{Margon}, B. 1984, \bibinfo{title}{{Observations of SS 433},} Annual Review of Astronomy and Astrophysics, 22, 507, \dodoi{10.1146/annurev.aa.22.090184.002451}

\bibitem[{J.~C. McKinney {et~al.}(2013)McKinney, Tchekhovskoy, \& Blandford}]{McKinney2013}
McKinney, J.~C., Tchekhovskoy, A., \& Blandford, R.~D. 2013, \bibinfo{title}{{Alignment of magnetized accretion disks and relativistic jets with spinning black holes},} Science, 339, 49, \dodoi{10.1126/science.1230811}

\bibitem[{J.~C.~A. {Miller-Jones} {et~al.}(2019){Miller-Jones}, {Tetarenko}, {Sivakoff}, {Middleton}, {Altamirano}, {Anderson}, {Belloni}, {Fender}, {Jonker}, {K{\"o}rding}, {Krimm}, {Maitra}, {Markoff}, {Migliari}, {Mooley}, {Rupen}, {Russell}, {Russell}, {Sarazin}, {Soria}, \& {Tudose}}]{2019Natur.569..374M}
{Miller-Jones}, J. C.~A., {Tetarenko}, A.~J., {Sivakoff}, G.~R., {et~al.} 2019, \bibinfo{title}{{A rapidly changing jet orientation in the stellar-mass black-hole system V404 Cygni},} \nat, 569, 374, \dodoi{10.1038/s41586-019-1152-0}

\bibitem[{S.~E. {Motta} {et~al.}(2014){Motta}, {Belloni}, {Stella}, {Mu{\~n}oz-Darias}, \& {Fender}}]{2014MNRAS.437.2554M}
{Motta}, S.~E., {Belloni}, T.~M., {Stella}, L., {Mu{\~n}oz-Darias}, T., \& {Fender}, R. 2014, \bibinfo{title}{{Precise mass and spin measurements for a stellar-mass black hole through X-ray timing: the case of GRO J1655-40},} \mnras, 437, 2554, \dodoi{10.1093/mnras/stt2068}

\bibitem[{R. {Narayan} \& E. {Quataert}(2005){Narayan} \& {Quataert}}]{Narayan2003}
{Narayan}, R., \& {Quataert}, E. 2005, \bibinfo{title}{{Black Hole Accretion},} Science, 307, 77, \dodoi{10.1126/science.1105746}

\bibitem[{D.~R. {Pasham} {et~al.}(2024){Pasham}, {Zaja{\v{c}}ek}, {Nixon}, {Coughlin}, {{\'S}niegowska}, {Janiuk}, {Czerny}, {Wevers}, {Guolo}, {Ajay}, \& {Loewenstein}}]{2024Natur.630..325P}
{Pasham}, D.~R., {Zaja{\v{c}}ek}, M., {Nixon}, C.~J., {et~al.} 2024, \bibinfo{title}{{Lense-Thirring precession after a supermassive black hole disrupts a star},} \nat, 630, 325, \dodoi{10.1038/s41586-024-07433-w}

\bibitem[{P. {Polko} \& J.~C. {McKinney}(2017){Polko} \& {McKinney}}]{2017MNRAS.464.2660P}
{Polko}, P., \& {McKinney}, J.~C. 2017, \bibinfo{title}{{Electromagnetic versus Lense-Thirring alignment of black hole accretion discs},} Monthly Notices of the Royal Astronomical Society, 464, 2660, \dodoi{10.1093/mnras/stw1875}

\bibitem[{O. Porth {et~al.}(2017)Porth, Olivares, Mizuno, Younsi, Rezzolla, Moscibrodzka, Falcke, \& Kramer}]{Porth2017}
Porth, O., Olivares, H., Mizuno, Y., {et~al.} 2017, \bibinfo{title}{{The black hole accretion code},} Computational Astrophysics and Cosmology, 4, 1, \dodoi{10.1186/s40668-017-0020-2}

\bibitem[{O. {Porth} {et~al.}(2019){Porth}, {Chatterjee}, {Narayan}, {Gammie}, {Mizuno}, {Anninos}, {Baker}, {Bugli}, {Chan}, {Davelaar}, {Del Zanna}, {Etienne}, {Fragile}, {Kelly}, {Liska}, {Markoff}, {McKinney}, {Mishra}, {Noble}, {Olivares}, {Prather}, {Rezzolla}, {Ryan}, {Stone}, {Tomei}, {White}, {Younsi}, {Akiyama}, {Alberdi}, {Alef}, {Asada}, {Azulay}, {Baczko}, {Ball}, {Balokovi{\'c}}, {Barrett}, {Bintley}, {Blackburn}, {Boland}, {Bouman}, {Bower}, {Bremer}, {Brinkerink}, {Brissenden}, {Britzen}, {Broderick}, {Broguiere}, {Bronzwaer}, {Byun}, {Carlstrom}, {Chael}, {Chatterjee}, {Chen}, {Chen}, {Cho}, {Christian}, {Conway}, {Cordes}, {Geoffrey}, {Crew}, {Cui}, {De Laurentis}, {Deane}, {Dempsey}, {Desvignes}, {Doeleman}, {Eatough}, {Falcke}, {Fish}, {Fomalont}, {Fraga-Encinas}, {Freeman}, {Friberg}, {Fromm}, {G{\'o}mez}, {Galison}, {Garc{\'\i}a}, {Gentaz}, {Georgiev}, {Goddi}, {Gold}, {Gu}, {Gurwell}, {Hada}, {Hecht}, {Hesper}, {Ho}, {Ho}, {Honma}, {Huang}, {Huang}, {Hughes}, {Ikeda}, {Inoue},
  {Issaoun}, {James}, {Jannuzi}, {Janssen}, {Jeter}, {Jiang}, {Johnson}, {Jorstad}, {Jung}, {Karami}, {Karuppusamy}, {Kawashima}, {Keating}, {Kettenis}, {Kim}, {Kim}, {Kim}, {Kino}, {Koay}, {Patrick}, {Koch}, {Koyama}, {Kramer}, {Kramer}, {Krichbaum}, {Kuo}, {Lauer}, {Lee}, {Li}, {Li}, {Lindqvist}, {Liu}, {Liuzzo}, {Lo}, {Lobanov}, {Loinard}, {Lonsdale}, {Lu}, {MacDonald}, {Mao}, {Marrone}, {Marscher}, {Mart{\'\i}-Vidal}, {Matsushita}, {Matthews}, {Medeiros}, {Menten}, {Mizuno}, {Moran}, {Moriyama}, {Moscibrodzka}, {M{\"u}ller}, {Nagai}, {Nagar}, {Nakamura}, {Narayanan}, {Natarajan}, {Neri}, {Ni}, {Noutsos}, {Okino}, {Oyama}, {{\"O}zel}, {Palumbo}, {Patel}, {Pen}, {Pesce}, {Pi{\'e}tu}, {Plambeck}, {PopStefanija}, {Preciado-L{\'o}pez}, {Psaltis}, {Pu}, {Ramakrishnan}, {Rao}, {Rawlings}, {Raymond}, {Ripperda}, {Roelofs}, {Rogers}, {Ros}, {Rose}, {Roshanineshat}, {Rottmann}, {Roy}, {Ruszczyk}, {Rygl}, {S{\'a}nchez}, {S{\'a}nchez-Arguelles}, {Sasada}, {Savolainen}, {Schloerb}, {Schuster}, {Shao}, {Shen}, {Small},
  {Sohn}, {SooHoo}, {Tazaki}, {Tiede}, {Tilanus}, {Titus}, {Toma}, {Torne}, {Trent}, \& {Trippe}}]{2019ApJS..243...26P}
{Porth}, O., {Chatterjee}, K., {Narayan}, R., {et~al.} 2019, \bibinfo{title}{{The Event Horizon General Relativistic Magnetohydrodynamic Code Comparison Project},} \apjs, 243, 26, \dodoi{10.3847/1538-4365/ab29fd}

\bibitem[{J. {Poutanen} {et~al.}(2022){Poutanen}, {Veledina}, {Berdyugin}, {Berdyugina}, {Jermak}, {Jonker}, {Kajava}, {Kosenkov}, {Kravtsov}, {Piirola}, {Shrestha}, {Perez Torres}, \& {Tsygankov}}]{2022Sci...375..874P}
{Poutanen}, J., {Veledina}, A., {Berdyugin}, A.~V., {et~al.} 2022, \bibinfo{title}{{Black hole spin{\textendash}orbit misalignment in the x-ray binary MAXI J1820+070},} Science, 375, 874, \dodoi{10.1126/science.abl4679}

\bibitem[{B. {Prather} {et~al.}(2021){Prather}, {Wong}, {Dhruv}, {Ryan}, {Dolence}, {Ressler}, \& {Gammie}}]{2021JOSS....6.3336P}
{Prather}, B., {Wong}, G., {Dhruv}, V., {et~al.} 2021, \bibinfo{title}{{iharm3D: Vectorized General Relativistic Magnetohydrodynamics},} The Journal of Open Source Software, 6, 3336, \dodoi{10.21105/joss.03336}

\bibitem[{B.~S. {Prather}(2024){Prather}}]{2024arXiv240801361P}
{Prather}, B.~S. 2024, \bibinfo{title}{{KHARMA: Flexible, Portable Performance for GRMHD},} arXiv e-prints, arXiv:2408.01361, \dodoi{10.48550/arXiv.2408.01361}

\bibitem[{S.~M. {Ressler} {et~al.}(2021){Ressler}, {Quataert}, {White}, \& {Blaes}}]{2021MNRAS.504.6076R}
{Ressler}, S.~M., {Quataert}, E., {White}, C.~J., \& {Blaes}, O. 2021, \bibinfo{title}{{Magnetically modified spherical accretion in GRMHD: reconnection-driven convection and jet propagation},} Monthly Notices of the Royal Astronomical Society, 504, 6076, \dodoi{10.1093/mnras/stab311}

\bibitem[{C.~S. {Reynolds}(2021){Reynolds}}]{2021ARA&A..59..117R}
{Reynolds}, C.~S. 2021, \bibinfo{title}{{Observational Constraints on Black Hole Spin},} \araa, 59, 117, \dodoi{10.1146/annurev-astro-112420-035022}

\bibitem[{P.~A.~G. {Scheuer} \& R. {Feiler}(1996){Scheuer} \& {Feiler}}]{1996MNRAS.282..291S}
{Scheuer}, P.~A.~G., \& {Feiler}, R. 1996, \bibinfo{title}{{The realignment of a black hole misaligned with its accretion disc},} \mnras, 282, 291, \dodoi{10.1093/mnras/282.1.291}

\bibitem[{D.~M. {Smith} {et~al.}(2002){Smith}, {Heindl}, \& {Swank}}]{2002ApJ...578L.129S}
{Smith}, D.~M., {Heindl}, W.~A., \& {Swank}, J.~H. 2002, \bibinfo{title}{{Orbital and Superorbital Periods of 1E 1740.7-2942 and GRS 1758-258},} \apjl, 578, L129, \dodoi{10.1086/344701}

\bibitem[{K.~A. Sorathia {et~al.}(2012)Sorathia, Reynolds, Stone, \& Beckwith}]{Sorathia2012}
Sorathia, K.~A., Reynolds, C.~S., Stone, J.~M., \& Beckwith, K. 2012, \bibinfo{title}{{Global simulations of accretion disks. I. Convergence and comparisons with local models},} Astrophysical Journal, 749, \dodoi{10.1088/0004-637X/749/2/189}

\bibitem[{J.~F. {Steiner} {et~al.}(2011){Steiner}, {Reis}, {McClintock}, {Narayan}, {Remillard}, {Orosz}, {Gou}, {Fabian}, \& {Torres}}]{2011MNRAS.416..941S}
{Steiner}, J.~F., {Reis}, R.~C., {McClintock}, J.~E., {et~al.} 2011, \bibinfo{title}{{The spin of the black hole microquasar XTE J1550-564 via the continuum-fitting and Fe-line methods},} \mnras, 416, 941, \dodoi{10.1111/j.1365-2966.2011.19089.x}

\bibitem[{N. {Stone} \& A. {Loeb}(2012){Stone} \& {Loeb}}]{2012PhRvL.108f1302S}
{Stone}, N., \& {Loeb}, A. 2012, \bibinfo{title}{{Observing Lense-Thirring Precession in Tidal Disruption Flares},} \prl, 108, 061302, \dodoi{10.1103/PhysRevLett.108.061302}

\bibitem[{A. {Tchekhovskoy} {et~al.}(2014){Tchekhovskoy}, {Metzger}, {Giannios}, \& {Kelley}}]{2014MNRAS.437.2744T}
{Tchekhovskoy}, A., {Metzger}, B.~D., {Giannios}, D., \& {Kelley}, L.~Z. 2014, \bibinfo{title}{{Swift J1644+57 gone MAD: the case for dynamically important magnetic flux threading the black hole in a jetted tidal disruption event},} \mnras, 437, 2744, \dodoi{10.1093/mnras/stt2085}

\bibitem[{A. {Tchekhovskoy} {et~al.}(2011){Tchekhovskoy}, {Narayan}, \& {McKinney}}]{2011MNRAS.418L..79T}
{Tchekhovskoy}, A., {Narayan}, R., \& {McKinney}, J.~C. 2011, \bibinfo{title}{{Efficient generation of jets from magnetically arrested accretion on a rapidly spinning black hole},} Monthly Notices of the Royal Astronomical Society, 418, L79, \dodoi{10.1111/j.1745-3933.2011.01147.x}

\bibitem[{H. {Thirring}(1918){Thirring}}]{1918PhyZ...19...33T}
{Thirring}, H. 1918, \bibinfo{title}{{{\"U}ber die Wirkung rotierender ferner Massen in der Einsteinschen Gravitationstheorie.},} Physikalische Zeitschrift, 19, 33

\bibitem[{C.~J. {White} {et~al.}(2019){White}, {Quataert}, \& {Blaes}}]{2019ApJ...878...51W}
{White}, C.~J., {Quataert}, E., \& {Blaes}, O. 2019, \bibinfo{title}{{Tilted Disks around Black Holes: A Numerical Parameter Survey for Spin and Inclination Angle},} The Astrophysical Journal, 878, 51, \dodoi{10.3847/1538-4357/ab089e}

\bibitem[{G.~N. {Wong} {et~al.}(2022){Wong}, {Prather}, {Dhruv}, {Ryan}, {Mo{\'s}cibrodzka}, {Chan}, {Joshi}, {Yarza}, {Ricarte}, {Shiokawa}, {Dolence}, {Noble}, {McKinney}, \& {Gammie}}]{2022ApJS..259...64W}
{Wong}, G.~N., {Prather}, B.~S., {Dhruv}, V., {et~al.} 2022, \bibinfo{title}{{PATOKA: Simulating Electromagnetic Observables of Black Hole Accretion},} The Astrophysical Journals, 259, 64, \dodoi{10.3847/1538-4365/ac582e}

\bibitem[{F. Yuan {et~al.}(2022)Yuan, Wang, \& Yang}]{Yuan2022}
Yuan, F., Wang, H., \& Yang, H. 2022, \bibinfo{title}{{The Accretion Flow in M87 is Really MAD},} The Astrophysical Journal, 924, 124, \dodoi{10.3847/1538-4357/ac4714}

\bibitem[{J.~J. {Zanazzi} \& D. {Lai}(2019){Zanazzi} \& {Lai}}]{2019MNRAS.487.4965Z}
{Zanazzi}, J.~J., \& {Lai}, D. 2019, \bibinfo{title}{{Tidal disruption event discs around supermassive black holes: disc warp and inclination evolution},} \mnras, 487, 4965, \dodoi{10.1093/mnras/stz1610}

\bibitem[{A.~A. {Zdziarski} {et~al.}(2024){Zdziarski}, {Chand}, {Banerjee}, {Szanecki}, {Janiuk}, {Lubi{\'n}ski}, {Nied{\'z}wiecki}, {Dewangan}, \& {Misra}}]{2024ApJ...967L...9Z}
{Zdziarski}, A.~A., {Chand}, S., {Banerjee}, S., {et~al.} 2024, \bibinfo{title}{{What Is the Black Hole Spin in Cyg X-1?},} \apjl, 967, L9, \dodoi{10.3847/2041-8213/ad43ed}

\end{thebibliography}
\bibliographystyle{aasjournalv7}



\end{document}